%% file: Morokuma2017KISS14k_arxiv201707a.tex
\documentclass[]{pasj01}
\begin{document} 
\Received{}%{yyyy/mm/dd}
\Accepted{}%{yyyy/mm/dd}
%\Published{yyyy/mm/dd}

\title{
%Title of Your Paper
OISTER Optical and Near-Infrared Monitoring Observations of a Peculiar Radio-Loud Active Galactic Nucleus 
SDSS~J110006.07+442144.3
}

%%% begin:list of authors
%% Do NOT capitalize all letters in "textsc".
%\author{A-Firstname \textsc{A-Familyname}\altaffilmark{1}%
%\thanks{Example: Present Address is xxxxxxxxxx}}
%\altaffiltext{1}{A-Address of Institute}
%\email{aaaaa@xxx.xxx.xx.xx}
%
%\author{B-Firstname \textsc{B-Familyname},\altaffilmark{2}}
%\altaffiltext{2}{B-Address of Institute}
%\email{bbbbb@xxx.xxx.xx.xx}
%
%\author{C-Firstname \textsc{C-Familyname}\altaffilmark{3}}
%\altaffiltext{3}{C-Address of Institute}
%\email{ccccc@xxx.xxx.xx.xx}
%
\author{Tomoki \textsc{Morokuma}\altaffilmark{1}}
\email{tmorokuma@ioa.s.u-tokyo.ac.jp}
\author{Masaomi \textsc{Tanaka}\altaffilmark{2}}
\author{Yasuyuki T. \textsc{Tanaka}\altaffilmark{3}}
\author{Ryosuke \textsc{Itoh}\altaffilmark{4,5}}
\author{Nozomu \textsc{Tominaga}\altaffilmark{6,7}}
\author{Poshak \textsc{Gandhi}\altaffilmark{8}}
\author{Elena \textsc{Pian}\altaffilmark{9,10}}
\author{Paolo \textsc{Mazzali}\altaffilmark{11,12}}
\author{Kouji \textsc{Ohta}\altaffilmark{13}}
\author{Emiko \textsc{Matsumoto}\altaffilmark{6}}
\author{Takumi \textsc{Shibata}\altaffilmark{6}}
\author{Hinako \textsc{Akimoto}\altaffilmark{14}}
\author{Hiroshi \textsc{Akitaya}\altaffilmark{15,3}}
\author{Gamal B. \textsc{Ali}\altaffilmark{16}}
\author{Tsutomu \textsc{Aoki}\altaffilmark{17}}
\author{Mamoru \textsc{Doi}\altaffilmark{1,18}}
\author{Nana \textsc{Ebisuda}\altaffilmark{5}}
\author{Ahmed \textsc{Essam}\altaffilmark{16}}
\author{Kenta \textsc{Fujisawa}\altaffilmark{19}}
\author{Hideo \textsc{Fukushima}\altaffilmark{2}}
\author{Shuhei \textsc{Goda}\altaffilmark{20}}
\author{Yuya \textsc{Gouda}\altaffilmark{20}}
\author{Hidekazu \textsc{Hanayama}\altaffilmark{21}}
\author{Yasuhito \textsc{Hashiba}\altaffilmark{1,36}}
\author{Osamu \textsc{Hashimoto}\altaffilmark{22}}
\author{Kenzo \textsc{Hayashida}\altaffilmark{23}}
\author{Yuichiro \textsc{Hiratsuka}\altaffilmark{24}}
\author{Satoshi \textsc{Honda}\altaffilmark{14}}
\author{Masataka \textsc{Imai}\altaffilmark{20}}
\author{Kanichiro \textsc{Inoue}\altaffilmark{23}}
\author{Michiko \textsc{Ishibashi}\altaffilmark{24}}
\author{Ikuru \textsc{Iwata}\altaffilmark{25}}
\author{Hideyuki \textsc{Izumiura}\altaffilmark{26}}
\author{Yuka \textsc{Kanda}\altaffilmark{5}}
\author{Miho \textsc{Kawabata}\altaffilmark{5}}
\author{Kenji \textsc{Kawaguchi}\altaffilmark{5}}
\author{Nobuyuki \textsc{Kawai}\altaffilmark{4}}
\author{Mitsuru \textsc{Kokubo}\altaffilmark{27,36,1}}
\author{Daisuke \textsc{Kuroda}\altaffilmark{26}}
\author{Hiroyuki \textsc{Maehara}\altaffilmark{26,17}}
\author{Hiroyuki \textsc{Mito}\altaffilmark{17}}
\author{Kazuma \textsc{Mitsuda}\altaffilmark{1}}
\author{Ryota \textsc{Miyagawa}\altaffilmark{24}}
\author{Takeshi \textsc{Miyaji}\altaffilmark{21}}
\author{Yusuke \textsc{Miyamoto}\altaffilmark{28,15}}
\author{Kumiko \textsc{Morihana}\altaffilmark{14}}
\author{Yuki \textsc{Moritani}\altaffilmark{7,3}}
\author{Kana \textsc{Morokuma-Matsui}\altaffilmark{29,2}}
\author{Kotone \textsc{Murakami}\altaffilmark{23}}
\author{Katsuhiro L. \textsc{Murata}\altaffilmark{30}}
\author{Takahiro \textsc{Nagayama}\altaffilmark{23}}
\author{Kazuki \textsc{Nakamura}\altaffilmark{24}}
\author{Tatsuya \textsc{Nakaoka}\altaffilmark{5}}
\author{Kotaro \textsc{Niinuma}\altaffilmark{31}}
\author{Takafumi \textsc{Nishimori}\altaffilmark{23}}
\author{Daisaku \textsc{Nogami}\altaffilmark{13}}
\author{Yumiko \textsc{Oasa}\altaffilmark{24,32}}
\author{Tatsunori \textsc{Oda}\altaffilmark{24}}
\author{Tomohito \textsc{Ohshima}\altaffilmark{14}}
\author{Yoshihiko \textsc{Saito}\altaffilmark{4}}
\author{Shuichiro \textsc{Sakata}\altaffilmark{23}}
\author{Shigeyuki \textsc{Sako}\altaffilmark{1}}
\author{Yuki \textsc{Sarugaku}\altaffilmark{17}}
\author{Satoko \textsc{Sawada-Satoh}\altaffilmark{15}}
\author{Genta \textsc{Seino}\altaffilmark{24}}
\author{Kazuo \textsc{Sorai}\altaffilmark{33}}
\author{Takao \textsc{Soyano}\altaffilmark{17}}
\author{Francesco \textsc{Taddia}\altaffilmark{34}}
\author{Jun \textsc{Takahashi}\altaffilmark{14}}
\author{Yuhei \textsc{Takagi}\altaffilmark{2,14}}
\author{Katsutoshi \textsc{Takaki}\altaffilmark{5}}
\author{Koji \textsc{Takata}\altaffilmark{5}}
\author{Ken'ichi \textsc{Tarusawa}\altaffilmark{17}}
\author{Makoto \textsc{Uemura}\altaffilmark{3}}
\author{Takahiro \textsc{Ui}\altaffilmark{5}}
\author{Riku \textsc{Urago}\altaffilmark{23}}
\author{Kazutoshi \textsc{Ushioda}\altaffilmark{24}}
\author{Jun-ichi \textsc{Watanabe}\altaffilmark{2}}
\author{Makoto \textsc{Watanabe}\altaffilmark{35,20}}
\author{Satoshi \textsc{Yamashita}\altaffilmark{23}}
\author{Kenshi \textsc{Yanagisawa}\altaffilmark{26}}
\author{Yoshinori \textsc{Yonekura}\altaffilmark{15}}
\author{Michitoshi \textsc{Yoshida}\altaffilmark{25,3}}
%%% Others
%\author{ \textsc{}\altaffilmark{}}
%
%on behalf of OISTER and JVN
\altaffiltext{1}{Institute of Astronomy, Graduate School of Science, University of Tokyo, 2-21-1, Osawa, Mitaka, Tokyo 181-0015, Japan}
\altaffiltext{2}{National Astronomical Observatory of Japan, National Institutes of Natural Sciences, 2-21-1, Osawa, Mitaka, Tokyo 181-8588, Japan}
\altaffiltext{3}{Hiroshima Astrophysical Science Center, Hiroshima University, Higashi-Hiroshima, Hiroshima 739-8526, Japan}
\altaffiltext{4}{Department of Physics, Tokyo Institute of Technology, 2-12-1 Ookayama, Meguro-ku, Tokyo 152-8551, Japan}
\altaffiltext{5}{Department of Physical Science, Hiroshima University, Kagamiyama 1-3-1, Higashi-Hiroshima 739-8526, Japan}
\altaffiltext{6}{Department of Physics, Faculty of Science and Engineering, Konan University, 8-9-1 Okamoto, Kobe, Hyogo 658-8501, Japan}
\altaffiltext{7}{Kavli Institute for the Physics and Mathematics of the Universe (WPI), The University of Tokyo, 5-1-5 Kashiwanoha, Kashiwa, Chiba 277-8583, Japan}
\altaffiltext{8}{Department of Physics and Astronomy, University of Southampton, Highfield, Southampton SO17 1BJ, UK}
\altaffiltext{9}{Institute of Space Astrophysics and Cosmic Physics, via P. Gobetti 101, I-40129 Bologna, Italy}
\altaffiltext{10}{Scuola Normale Superiore, Piazza dei Cavalieri 7, I-56126 Pisa, Italy}
\altaffiltext{11}{Astrophysics Research Institute, Liverpool John Moores University, IC2, Liverpool Science Park, 146 Brownlow Hill, Liverpool L3 5RF, UK}
\altaffiltext{12}{Max-Planck-Institut fr Astrophysik, Karl-Schwarzschild-Str. 1, D-85748 Garching, Germany}
\altaffiltext{13}{Department of Astronomy, Graduate School of Science, Kyoto University, Sakyo-ku, Kyoto 606-8502, Japan}
\altaffiltext{14}{Nishi-Harima Astronomical Observatory, Center for Astronomy, University of Hyogo, 407-2 Nishigaichi, Sayo, Hyogo 679-5313, Japan}
\altaffiltext{15}{Center for Astronomy, Ibaraki University, 2-1-1 Bunkyo, Mito, Ibaraki 310-8512, Japan}
\altaffiltext{16}{National Research Institute of Astronomy and Geophysics, 11421 Helwn, Cairo, Egypt}
\altaffiltext{17}{Kiso Observatory, Institute of Astronomy, School of Science, The University of Tokyo 10762-30, Mitake, Kiso-machi, Kiso-gun, Nagano 397-0101, Japan}
\altaffiltext{18}{Research Center for the Early Universe, Graduate School of Science, The University of Tokyo, 7-3-1 Hongo, Bunkyo-ku, Tokyo 113-0033, Japan}
\altaffiltext{19}{The Research Institute of Time Studies, Yamaguchi University, 1677-1 Yoshida, Yamaguchi, Yamaguchi 753-8511, Japan}
\altaffiltext{20}{Department of Cosmosciences, Hokkaido University, Kita 10 Nishi 8, Kita-ku, Sapporo, Hokkaido 060-0810, Japan}
\altaffiltext{21}{Ishigakijima Astronomical Observatory, National Astronomical Observatory of Japan, National Institutes of Natural Sciences, Ishigaki, Okinawa 907-0024, Japan}
\altaffiltext{22}{Gunma Astronomical Observatory, Takayama, Gunma 377-0702, Japan}
\altaffiltext{23}{Graduate School of Science and Engineering, Kagoshima University, 1-21-35 Korimoto, Kagoshima 890-0065, Japan}
\altaffiltext{24}{Faculty of Education, Saitama University, 255 Shimo-Okubo, Sakura, Saitama, 338-8570, Japan}
\altaffiltext{25}{Subaru Telescope, National Astronomical Observatory of Japan, National Institutes of Natural Sciences, 650 North A'ohoku Place, Hilo, HI 96720, USA}
\altaffiltext{26}{Okayama Astrophysical Observatory, National Astronomical Observatory of Japan, National Institutes of Natural Sciences, 3037-5 Honjo, Kamogata, Asakuchi, Okayama 719-0232, Japan}
\altaffiltext{27}{Astronomical Institute, Tohoku University, Aramaki, Aoba-ku, Sendai, 980-8578, Japan}
\altaffiltext{28}{Nobeyama Radio Observatory, National Astronomical Observatory of Japan, National Institutes of Natural Sciences, 462-2 Nobeyama, Minamimaki, Nagano 384-1305, Japan}
\altaffiltext{29}{The Institute of Space and Astronautical Science, Japan Aerospace Exploration Agency, 3-1-1 Yoshinodai, Chuo-ku, Sagamihara, Kanagawa 252-5210, Japan}
\altaffiltext{30}{Department of Astrophysics, Nagoya University, Chikusa-ku, Nagoya 464-8602, Japan}
\altaffiltext{31}{Graduate School of Sciences and Technology for Innovation, Yamaguchi University, 1677-1 Yoshida, Yamaguchi, Yamaguchi 753-8512, Japan}
\altaffiltext{32}{Department of Physics, Graduate School of Science and Engineering, Saitama University, 255 Shimo-Okubo, Sakura, Saitama, 338-8570, Japan}
\altaffiltext{33}{Department of Physics, Faculty of Science, Hokkaido University, Kita-ku, Sapporo 060-0810, Japan}
\altaffiltext{34}{The Oskar Klein Centre, Department of Astronomy, Stockholm University, AlbaNova, SE-10691 Stockholm, Sweden}
\altaffiltext{35}{Department of Applied Physics, Okayama University of Science, 1-1 Ridai-cho, Kita-ku, Okayama, Okayama 700-0005, Japan}
\altaffiltext{36}{JSPS Fellow}
%\email{tmorokuma@ioa.s.u-tokyo.ac.jp}
%%% end:list of authors
%% `\KeyWords{}' always has to be placed before `\maketitle'.
 %Do NOT move this preamble from here!

\maketitle

\begin{abstract}
%Please read ``IMPORTANT NOTICE'' carefully before preparing a manuscript. 
We present monitoring campaign observations at optical and near-infrared (NIR) 
wavelengths for a radio-loud active galactic nucleus (AGN) at $z=0.840$, SDSS~J110006.07+442144.3 (hereafter, J1100+4421), 
which was identified during a flare phase in late February, 2014. 
The campaigns consist of three intensive observing runs from the discovery to March, 2015, mostly within the scheme of the OISTER collaboration. 
Optical-NIR light curves and simultaneous spectral energy distributions (SEDs) are obtained.
Our measurements show the strongest brightening in March, 2015.
We found that the optical-NIR SEDs of J1100+4421 show an almost steady shape despite the large and rapid intranight variability. 
This constant SED shape is confirmed to extend to $\sim5~\mu$m in the observed frame using the archival WISE data. 
Given the lack of absorption lines and the steep power-law spectrum of $\alpha_{\nu}\sim-1.4$, where $f_{\nu}\propto\nu^{\alpha_{\nu}}$, 
synchrotron radiation by a relativistic jet with no or small contributions from the host galaxy and the accretion disk seems most plausible as an optical-NIR emission mechanism. 
The steep optical-NIR spectral shape and the large amplitude of variability are consistent with this object being a low $\nu_{\rm{peak}}$ jet-dominated AGN.
In addition, sub-arcsec resolution optical imaging data taken with Subaru Hyper Suprime-Cam 
does not show a clear extended component and the spatial scales are significantly smaller than the large extensions detected at radio wavelengths. 
The optical spectrum of a possible faint companion galaxy does not show any emission lines at the same redshift and 
hence a merging hypothesis for this AGN-related activity is not supported by our observations. 
\end{abstract}

%%%%%%%%%%%%%%%%%%%%%%
%%%%%%%%%%%%%%%%%%%%%%
%%%%%%%%%%%%%%%%%%%%%%
\section{Introduction}\label{sec:sec_intro}
%%%%%%%%%%%%%%%%%%%%%%
%%%%%%%%%%%%%%%%%%%%%%
%%%%%%%%%%%%%%%%%%%%%%

Since the detection of $\gamma$-ray emission from narrow-line Seyfert 1 galaxies (NLS1s; \cite{abdo2009}), 
which are one of the low-mass populations of active galactic nuclei (AGN), 
the nature of systems with relativistic jets 
but lower masses than blazars and radio galaxies has attracted much attention 
from researchers in fields such as AGN, relativistic phenomena, and galaxies. 
These $\gamma$-ray-loud NLS1s have in general smaller black hole (BH) masses 
of $M_{\rm{BH}}\sim10^{6-8}$~M$_\odot$ \citep{abdo2009} for the first four $\gamma$-ray-loud NLS1s 
than blazars with larger BHs of $<M_{\rm{BH}}>=2.8\times10^{8}$~M$_\odot$\citep{pian2005}, 
even though such strong jet activities are thought to be generally associated with massive systems. 
Multi-wavelength observational studies on these populations,
including $\gamma$-ray \citep{abdo2009_nls1}, X-ray, high-resolution radio VLBI \citep{doi2012}, 
provide clues to the relationship between the central engines and jet formation mechanisms, 
AGN triggering mechanisms, 
and possibly BH growth (from lower to higher masses in quasars and elliptical galaxies). 
However, these observational studies have been done for a limited number of objects 
while statistical studies on NLS1s were recently done in several papers (e.g., \cite{foschini2015}). 

The rapid optical 
flare of the extragalactic object SDSS~J110006.07+442144.3 (J1100+4421, hereafter) 
was initially recognized on February 23, 2014 \citep{tanaka2014}, 
during the 1-hour-cadence Kiso Supernova Survey (KISS; \cite{morokuma2014}),
which utilizes the 105-cm Kiso Schmidt telescope and its optical wide-field imager, the Kiso Wide Field Camera (KWFC; \cite{sako2012}). 
Quick imaging and spectroscopic follow-up observations with the Faint Object Camera and Spectrograph (FOCAS; \cite{kashikawa2002}) 
on the Subaru 8.2-m telescope right after the flare detection 
indicated that the object was a narrow-line AGN at $z=0.840$ 
with a small BH mass of $M_{\rm{BH}}=1.0-1.5\times10^{7}$~M$_\odot$ 
and a sub-Eddington ratio of $L_{\rm{bol}}/L_{\rm{Edd}}\sim0.3$ \citep{tanaka2014}.
By combining these optical data with archival radio, infrared (WISE), $\gamma$-ray (Fermi) data 
and X-ray data taken via a ToO observation with X-ray Telescope~(XRT) on board the Swift satellite, 
\citet{tanaka2014} revealed the similarity of this object's SED to 
those of the radio/$\gamma$-ray-loud 
NLS1s, although the large [O$_{\rm{~III}}$]/H$\beta$ flux ratio does not satisfy the general criteria for NLS1s. 
Flux ratio between the [O$_{\rm{~III}}$] emission lines and radio emission is also 
consistent with those of radio-loud NLS1s \citep{berton2016}. 

To examine the origins of the emission mechanisms and variability of J1100+4421,
we carried out intensive monitoring observations 
under the auspices of the global telescope network known as Optical and Infrared Synergetic Telescopes for Education
and Research (OISTER\footnote{http://oister.oao.nao.ac.jp/}; \cite{sekiguchi2016}). 
%consisting of 14 telescopes with 21 instruments spread over the world, in Japan, South Africa, and Chile. 
OISTER is one of the best observational organizations to achieve 
continuous, long-term monitoring observations in optical and near-infrared wavelengths 
for relatively bright objects 
because of its multi-site observing facilities.

In this paper, 
we describe our imaging monitoring observations and additional spectroscopic observations in Section~\ref{sec:sec_obs}. 
Results of the data analyses for the observational data are shown in Section~\ref{sec:sec_results}. 
The implications of the observational results on the emission mechanisms and the host galaxy and environmental properties
are described in Section~\ref{sec:sec_discussions}. 
We summarize the content of the paper in Section~\ref{sec:sec_summary}. 

Cosmological parameters used in this paper are 
$\Omega_{M}=0.3, \Omega_{\Lambda}=0.7, H_{0}=70$~km~s$^{-1}$~Mpc$^{-1}$. 
All the observing times are specified in UT. 
All magnitudes in optical and near-infrared wavelengths are measured in AB system unless otherwise noted. 
Galactic extinctions \citep{schlafly2011}, 
$A_u=0.055, A_g=0.043, A_r=0.029, A_i=0.022, A_z=0.016, A_J=0.009, A_H=0.006$, and $A_{K_s}=0.004$~mag, 
are not corrected for 
the photometric values in Table~\ref{tab:tab_obs_niropt} or 
the spectra shown in Figure~\ref{fig:fig_spec} 
while they are corrected in calculating the SEDs and fitted power-law indices. 
Note that power-law indices of the spectral energy distribution (SED), $\alpha_{\nu}$, measured in this paper 
are defined as $f_{\nu}\propto\nu^{\alpha_{\nu}}$. 

%%%%%%%%%%%%%%%%%%%%%%
%%%%%%%%%%%%%%%%%%%%%%
%%%%%%%%%%%%%%%%%%%%%%
\section{Monitoring Campaign Observations and Data Reduction}\label{sec:sec_obs}
%%%%%%%%%%%%%%%%%%%%%%
%%%%%%%%%%%%%%%%%%%%%%
%%%%%%%%%%%%%%%%%%%%%%

%%%%%%%%%%%%%%%%%%%%%%%%%%%%
\begin{figure*}
 \begin{center}
  \includegraphics[width=120mm]{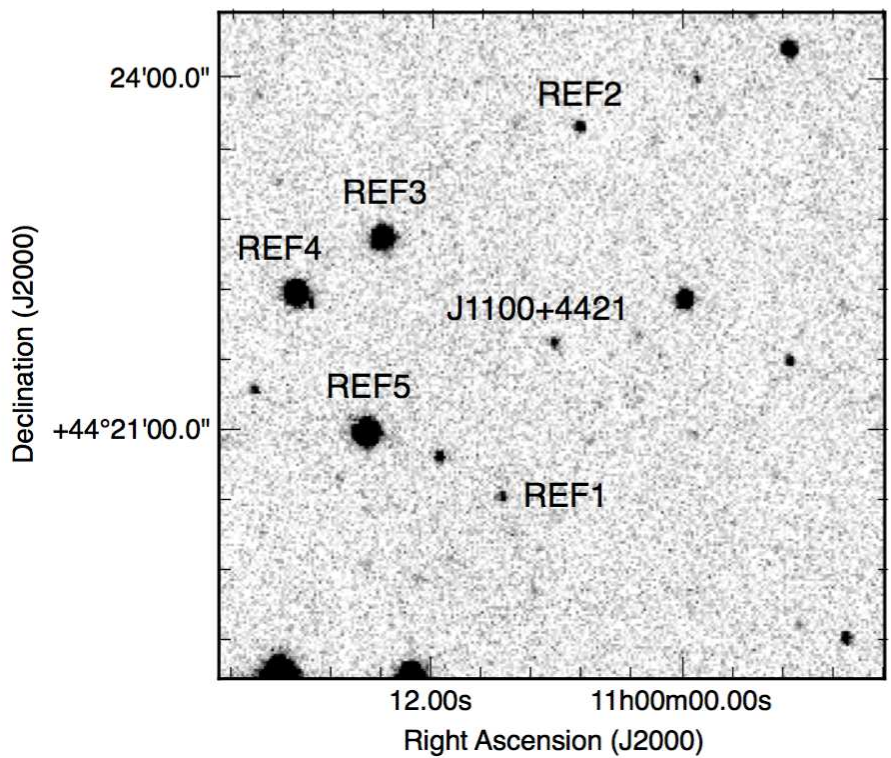}
 \end{center}
\caption{
The Kiso KWFC $g$-band image of the field at the discovery epoch. 
The central object in the figure is the target J1100+4421. 
The reference stars for the relative photometry are marked. 
North is up and east is left. The box size is 6~arcmin square. 
}\label{fig:fig_imgfield}
\end{figure*}
%%%%%%%%%%%%%%%%%%%%%%%%%%%%

%%%%%%%%%%%%%%%%%%%%%%
\subsection{Optical and Near-Infrared Imaging}\label{sec:sec_followup_niropt_data}
\subsubsection{Observations}\label{sec:sec_followup_niropt}
%%%%%%%%%%%%%%%%%%%%%%

We used 10 telescopes and 12 instruments in Japan among the OISTER 
collaboration to take most of the imaging data presented in this paper. 
Instruments used (from east to west in Japan) are 
a visible Multi-Spectral Imager (MSI; \cite{watanabe2012}) on the 1.6-m Pirka telescope, 
the FLI Micro Line Deep Depletion CCD camera on the 0.55-m Saitama university Common-use Research for Astronomy (SaCRA) telescope, 
MITSuME (\cite{kotani2005}; \cite{yatsu2007}; \cite{shimokawabe2008}) on the 0.5-m Akeno telescope, 
KWFC \citep{sako2012} on the 1.05-m Kiso Schmidt telescope, 
Line Imager and Slit Spectrograph (LISS; \cite{hashiba2014}) and 
Nishi-Harima Infrared Camera (NIC) on the 2.0-m Nayuta telescope, 
Kyoto Okayama Optical Low-dispersion Spectrograph (KOOLS; \cite{yoshida2005}) and 
ISLE \citep{yanagisawa2008} on the 1.88-m Okayama Astrophysical Observatory (OAO) telescope, 
MITSuME on the 0.5-m OAO telescope (\cite{kotani2005}; \cite{yanagisawa2010}), 
Hiroshima Optical and Near-InfraRed camera (HONIR; \cite{akitaya2014}; \cite{sakimoto2012}) on the 1.5-m Kanata telescope, 
a near-infrared camera on the 1.0-m Kagoshima telescope, 
and MITSuME on the 1.05-m Murikabushi telescope. 

In addition, we also took 
$B$-band images with the 1.88-m telescope at the Kottamia Astronomical Observatory (KAO) in Egypt on Feb. 9, 2015 and 
imaging data with Hyper Suprime-Cam (HSC; \cite{miyazaki2012}) on the Subaru 8.2-m telescope in Hawaii, 
in $g$-band on November 27, 2014 and May 24, 2015, and in $z$-band on Novermber 26, 2014, respectively. 
The plate scales and fields-of-view of these two instruments are 
0.304~arcsec$^{-1}$~pixel and 10~arcmin, and 
0.168~arcsec$^{-1}$~pixel and 1.5~degree in diameter of a circular shape field-of-view, respectively. 

The observing epochs fall into three intensive campaigns:
from February, 2014 to March, 2014, right after the discovery described in \citet{tanaka2014}, 
from October, 2014 to November, 2014, and in March 2015. 
In the second observing campaign, from Oct 2014 to Nov 2014, 
we obtained data at least once in multiple bands almost every night if the weather permitted. 
During the third observing epoch, 
optical imaging data with OAO188 KOOLS were taken 2-3 times per night over 4 continuous nights, at intervals of a few hours, if the weather permitted. 

The instruments used in this campaign nicely cover a wide range of wavelengths 
from optical to NIR, from $u$-band to $K_s$-band as shown in Table~\ref{tab:tab_obs_niropt}. 

%%%%%%%%%%%%%%%%%%%%%%
\subsubsection{Data Reduction}\label{sec:sec_datared}
%%%%%%%%%%%%%%%%%%%%%%

Basic data reduction for imaging data was done in a standard manner. 
Bias subtraction, overscan subtraction (if necessary), flat-fielding, sky subtraction, and astrometry procedures 
are done with a data reduction pipeline for each instrument. 
We combined all images taken within a night
to create one stacked image per day,
except for the discovery and quick follow-up data shown in \citet{tanaka2014} and 
the OAO KOOLS optical data in March 2015. 
The HSC data was reduced using hscPipe version 3.8.5, which was developed based on the LSST 
pipeline (\cite{ivezic2008}; \cite{axelrod2010}). 

For these reduced images, we performed forced photometry 
by fixing the coordinates of J1100+4421 and the reference stars (Table~\ref{tab:tab_refstars}) into consistent locations,
using the {\it Common-use Automatic Realtime Photometry} (CARP) pipeline \citep{saito2016}, 
which has been developed to conduct photometry for images taken by the OISTER collaboration,
The adopted aperture size was set to be three times the seeing FWHM in diameter. 
For images with $<3\sigma$ detections for J1100+4421, we set $3\sigma$ upper limits 
provided by the CARP pipeline. 

According to the fields-of-view and filters of the images, 
different reference stars may be used 
in different images,
although the same stars were used in a given filter as much as possible. 
Most of the data were calibrated via 
star \#2 in optical and the star \#3 in NIR. 
In the optical wavelengths, 
$ugriz$-band magnitudes of the reference stars are derived 
from SDSS Data Release~12~(DR12; \cite{alam2015}) 
and $BVRI$-band magnitudes are calculated from the SDSS magnitudes using the conversion equations shown in \citet{jester2005}. 
Magnitudes of the reference stars in NIR wavelengths are derived from the 2MASS database~\citep{skrutskie2006} 
and converted to those in the AB system as follows:
$J_{\rm{AB}}=J_{\rm{Vega}}+0.94$, 
$H_{\rm{AB}}=H_{\rm{Vega}}+1.38$, and 
$K_{s,\rm{AB}}=K_{s,\rm{Vega}}+1.86$ \citep{tokunaga2005}. 

The reference stars \#1, and \#2 have nearby stars separated by 
3.5~arcsec and 6.2~arcsec, respectively, 
which are at least $3.4$~mag fainter in optical wavelengths than the reference stars. 
Most of our imaging data were taken under seeing of a few arcsec and 
these nearby stars contaminate the flux measurements of the reference stars. 
However, the fraction of flux contributed by these nearby stars is 
estimated to be 
typically a factor of $\sim0.02$ 
by comparing the SDSS magnitudes of the stars 
and does not affect our conclusions. 

%%%%%%%%%%%%%%%%%%%%%%%%%%%
%%% reference star data %%%
%%%%%%%%%%%%%%%%%%%%%%%%%%%
\begin{table*}
  \tbl{SDSS and 2MASS photometry of the References Stars.}{%
  \begin{tabular}{llllllllllll}
      \hline
      Ref & RA & Dec & $u$ & $g$ & $r$ & $i$ & $z$ & $J$ & $H$ & $K_s$\\\hline
      \hline
	1	&	11:00:08.54	&	+44:20:25.95	&	23.81(0.68)	&	20.26(0.02)	&	18.80(0.01)	&	17.94(0.01)	&	17.51(0.01)	&	16.40(0.11)	&	15.69(0.12)	&	15.59(0.17)	\\ 
	2	&	11:00:04.87	&	+44:23:34.51	&	21.96(0.17)	&	19.67(0.01)	&	18.75(0.01)	&	18.39(0.01)	&	18.21(0.02)	& - & - & - \\ 
	3	&	11:00:14.17	&	+44:22:38.39	&	18.64(0.02)	&	16.15(0.00)	&	15.08(0.00)	&	14.66(0.00)	&	14.46(0.00)	&	13.36(0.02)	&	12.72(0.02)	&	12.60(0.02)	\\ 
	4	&	11:00:18.37	&	+44:22:09.50	&	17.65(0.01)	&	15.66(0.00)	&	14.97(0.00)	&	14.70(0.00)	&	14.60(0.00)	&	13.62(0.02)	&	13.15(0.02)	&	13.06(0.03)	\\ 
	5	&	11:00:14.97	&	+44:20:58.44	&	16.73(0.01)	&	14.93(0.00)	&	14.31(0.00)	&	14.59(0.00)	&	13.97(0.00)	&	13.01(0.02)	&	12.61(0.02)	&	12.49(0.02)	\\ 
	\hline
    \end{tabular}}\label{tab:tab_refstars}
\begin{tabnote}
$JHK$ magnitudes are shown in the Vega system, which are the raw values in the 2MASS catalog. 
Magnitude errors are shown in the parenthesis. 
\end{tabnote}
\end{table*}

In principle,
using different reference stars for relative photometry could provide different magnitudes for the target.
The spectral slopes of J1100+4421, as described in \S\ref{sec:sec_sed}, are comparable to those of cool, red stars. 
We checked for possible systematic differences in relative photometry with different stars of different spectral slopes (colors) 
and confirmed that the differences of the resulting magnitudes are smaller than 10\%,
even if we use a blue star as a reference star. 
Therefore, we conclude that the choice of the reference stars does not affect our conclusions, 
which require only about 10\% accuracy. 

As shown in Section \ref{sec:sec_hostgal}, 
there is a faint galaxy 2.7~arcsec away from the object, 
which also contaminates the flux of the object. 
The measured brightness ($g=25.32$~mag, $z=22.15$~mag) 
is roughly 10~times or more fainter than the object in most of the detected epochs 
although these factors depend on the brightness phase of the object. 
These are also small effects on our photometries and resultant discussion. 

%%%%%%%%%%%%%%%%%%%%%%%%
\subsection{Optical Spectroscopy}\label{sec:sec_followup_optspec_data}
\subsubsection{Observations}\label{sec:sec_followup_optspec}
%%%%%%%%%%%%%%%%%%%%%%%%

In addition to the Subaru FOCAS spectra obtained in February, 2014 \citep{tanaka2014}, 
we took a new long-slit spectrum of a 
900-sec exposure using the grating R400 and the order-sort filter OG515
with the Gemini Multi-Object Spectrographs (GMOS; \cite{hook2004}) 
on the Gemini-North 8.1-m telescope on December 11, 2015. 
The resulting wavelength range is 5150-9400\AA. 
The slit width was 1.0~arcsec and the spectral resolution was $R\sim940$. 
The slit was aligned with J1100+4421 and a nearby galaxy 
located about 2.7~arcsec to the south-east. 
Although the sky was clear, it was windy and then the seeing was bad and variable 
from 1.1~arcsec to 3.2~arcsec during the night.

We note that atmospheric dispersion corrector (ADC) is not available for GMOS 
while an ADC is equipped with the Cassegrain focus and available for FOCAS on the Subaru telescope. 
The resultant slit loss is estimated in some cases and available on the Gemini 
website\footnote{http://www.gemini.edu/sciops/instruments/gmos/itc-sensitivity-and-overheads/atmospheric-differential-refraction}. 
The position angle of the slit was 115.0~deg (east from north) and 
the parallactic angle was $\sim-140$~deg. 
Then, the difference of the angles was $\sim105$~deg causing 
the differential slit loss of $\sim20$\% over the observed wavelength range. 
On the other hand, the standard star used for flux calibration, Feige~110, was observed at the parallactic angle on the same night. 

%%%%%%%%%%%%%%%%%%%%%%%%%%%%%%%%%%%%%%%%
\begin{figure*}[htbp!]
\begin{center}
	\includegraphics[width=150mm,angle=0]{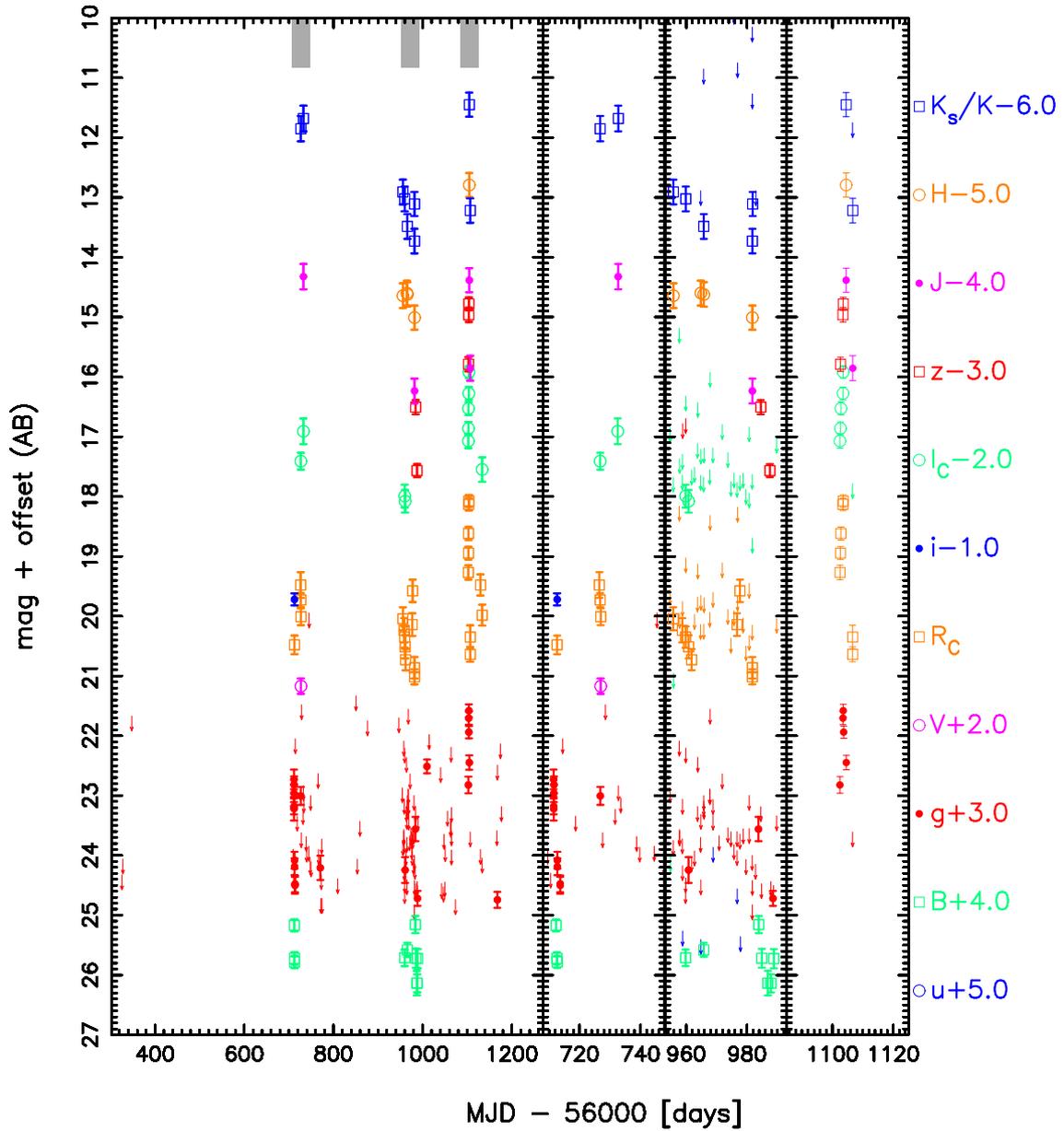}
\end{center}
\caption{
Light curves of SDSS~J110006.07+442144.3 in magnitude unit. 
$u$ in blue, 
$B$ in green, 
$g$ in red, 
$V$ in purple, 
$R_C$ in orange, 
$i$ in blue, 
$I_C$ in green, 
$z$ in red, 
$J$ in purple, 
$H$ in orange, and 
$K_s$ and $K$ in blue. 
{\it Left Panel}: Entire light curves of J1100+4421. Upper limits only in $g$-band are shown. 
Time ranges magnified in the right panels are shown in gray at the top of the panel. 
{\it Three Right Panels}: Light curves of J1100+4421 focused around the discovery epoch, the Oct-Nov 2014 campaign observing run, and the bright epoch in March 2015, respectively.
Upper limits in all the filters are shown.
Galactic extinctions are not corrected. 
\label{fig:fig_wholelc}}
\end{figure*}
%%%%%%%%%%%%%%%%%%%%%%%%%%%%%%%%%%%%%%%%

%%%%%%%%%%%%%%%%%%%%%%%%%%%%%%%%%%%%%%%%
\begin{figure*}[htbp!]
\begin{center}
	\includegraphics[angle=270,width=141mm]{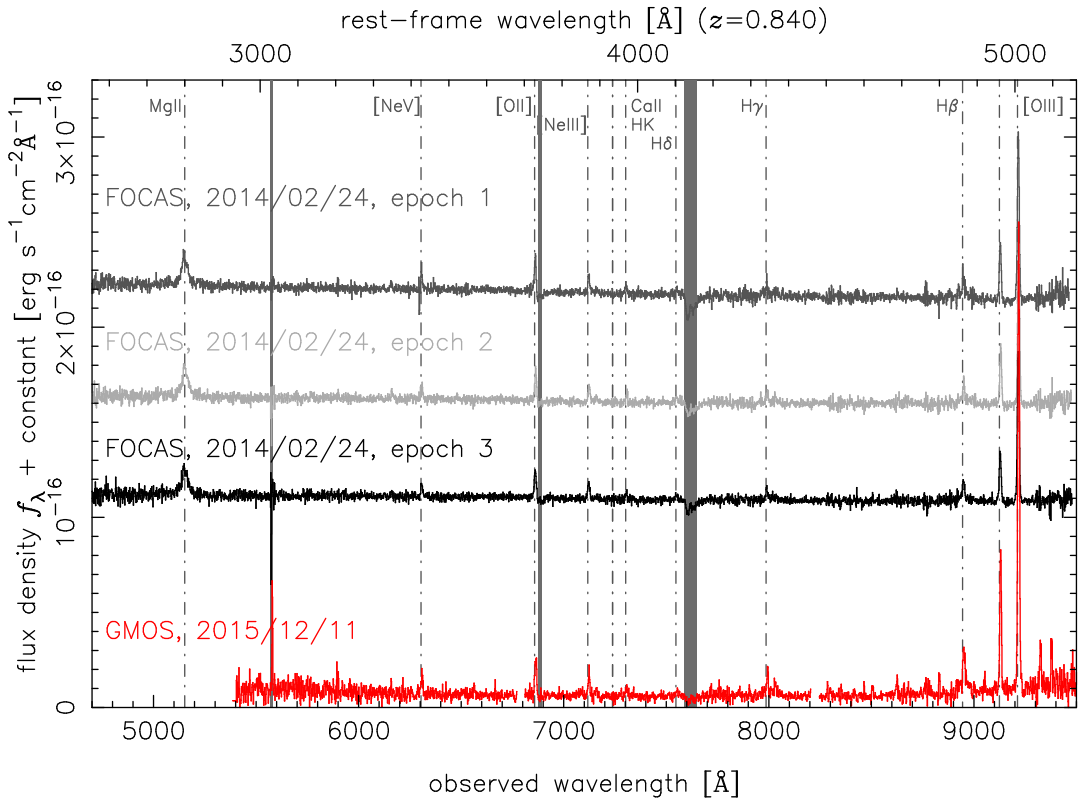}
\end{center}
\caption{
A comparison of optical spectra taken with 
Subaru FOCAS on February 24, 2014 (already shown in \cite{tanaka2014}) and 
Gemini-N GMOS newly obtained on December 11, 2015. 
The GMOS spectrum is separated into three wavelength ranges because the gaps 
between the adjacent GMOS CCDs.
Emission and absorption lines which could originate from an AGN and a galaxy are 
indicated in dot-dashed lines. Note that all the indicated lines are not detected. 
Strong telluric absorptions (O$_{2}$ A-band and B-band) and a strong OH airglow emission around 5,577\AA\
causing bad sky subtraction are shaded in dark gray. 
Galactic extinctions are not corrected. 
\label{fig:fig_spec}
}
\end{figure*}
%%%%%%%%%%%%%%%%%%%%%%%%%%%%%%%%%%%%%%%%

%%%%%%%%%%%%%%%%%%%%%%%%%%%%%%%%%%%%%%%%%%%%%%%%%%
\subsubsection{Data Reduction}\label{sec:sec_followup_optspec_red}
%%%%%%%%%%%%%%%%%%%%%%%%%%%%%%%%%%%%%%%%%%%%%%%%%%

The GMOS spectrum was reduced with the Gemini IRAF package. 
The reduction procedure was basically equivalent to that done for the FOCAS spectra \citep{tanaka2014}. 
Wavelength calibration was done using the CuAr lamp spectrum.
The obtained GMOS spectrum and the FOCAS spectra on February 23, 2014 \citep{tanaka2014} 
are shown in Figure~\ref{fig:fig_spec}. 
Note that the GMOS spectrum was obtained without the ADC and the relative flux calibration 
is not correct at $\sim20$\% level. 

%%%%%%%%%%%%%%%%%%%%%%
%%%%%%%%%%%%%%%%%%%%%%
%%%%%%%%%%%%%%%%%%%%%%
\section{Results}\label{sec:sec_results}
%%%%%%%%%%%%%%%%%%%%%%
%%%%%%%%%%%%%%%%%%%%%%
%%%%%%%%%%%%%%%%%%%%%%

%%%%%%%%%%%%%%%%%%%%%%
\subsection{Light Curves}\label{sec:sec_lc}
%%%%%%%%%%%%%%%%%%%%%%
The obtained light curves~(LCs) are shown in Figure~\ref{fig:fig_wholelc} and summarized in Table~2. 
Deep observations in which the object is detected with a high significance are limited 
because of the faintness of the object for 1-2m-class telescopes. 
The object, when detected, was brighter than the SDSS magnitudes 
measured for the $ugriz$-band images taken in 2003;
the SDSS magnitudes are the faintest among our data,
although the depths of most of our images are shallower than the SDSS magnitudes. 

As described in Section \ref{sec:sec_followup_niropt_data}, 
the observing data mostly consist of three campaigns. 
We below summarize the brightness changes separately in each of these three observing epochs. 

First, we briefly summarize the discovery with Kiso KWFC 
and $1$-month follow-up observations after the discovery, 
which were described in detail in \citet{tanaka2014}. 
The object showed a rapid increase in brightness
from non-detection ($g>20-21$~mag) to $19.73\pm0.13$~mag on February 23, 2014.
After that, the object quickly faded down to $21-22$~mag 
during the discovery night at Kiso 
and the next 24 hours at Mauna Kea, Hawaii. 
We monitored the object after this discovery over about 1~month with 
Kiso KWFC ($g$-band), 
Kottamia Observatory 1.88-m telescope ($V$-band), 
Akeno MITSuME ($g,R_C,I_C$-bands), 
OAO KOOLS ($g$-band), and 
Kanata HONIR ($R_C,J$-bands). 
The data sampling was sparse, but the object seemed to brighten a few times 
to a level similar to that of the discovery epoch. 
The peak magnitude in this observing epoch 
in blue optical bands was $V=19.17\pm0.13$~mag 
and that in NIR was $K_s=17.68\pm0.22$. 

In the second observing campaign, 
we did not detect the object 
on most of the nights during bright moon phases, placing only weak upper limits on its brightness at those times.
Compared to the first and third observing epochs, 
the object was fainter, reaching only to $R_C=19.58\pm0.19$~mag 
and $K_s=18.91\pm0.21$ 
even at its peak.

The object was at its brightest in this third observing epoch, 
up to $18.58\pm0.11$~mag in $g$-band (brighter than the discovery epoch by a factor of $\sim2.9$ in flux units) 
and $18.07\pm0.10$~mag in $R_C$-band. 
In addition, the object was also the brightest in NIR, $K_s=17.45\pm0.20$.  

%%%%%%%%%%%%%%%%%%%%%%%%%%%%%%%%%%%%%%%%
\subsection{Variability Time Scale}\label{sec:sec_varitime}
%%%%%%%%%%%%%%%%%%%%%%%%%%%%%%%%%%%%%%%%

Variability time scales, in general, provide useful information to constrain 
the size of an emitting region and the emission mechanism producing the variability. 
We adopt a doubling/halving time scale $\tau$ as an indicator of the variability, 
\begin{eqnarray}
	F(t)=F_0(t)2^{-(t-t_0)/\tau} 
\end{eqnarray}
following equation (1) in \citet{foschini2011}. 
As done in \citet{foschini2011} for flat-spectrum radio quasars at MeV energies, 
the time scale $\tau$ is calculated only for a combination of the measurements 
which show significant ($>3\sigma$) variability. 

The shortest halving time scale $\tau$ in the observed frame 
obtained in our study is $\tau=0.38\pm0.18$~days ($\sim9.2\pm4.5$~hours). 
This is derived from the data taken on MJD 56712, 
on the night of our quick follow-up observations with Subaru FOCAS. 
Because of our limited time sampling and depths, 
the real typical halving time scale could be shorter than the obtained value. 
Compared with previous works on variability time scales of blazars 
in UV, optical, and NIR wavelength regions \citep{urry1997},
in which optical flares over time scales from minutes to days are detected, 
the time scale of J1100+4421 is longer than most of those values,
even though the BH mass of our object is smaller than those in blazars by a few orders of magnitude. 
If we assume a Doppler boosting factor of $\delta=10$ for J1100+4421, similar to that of blazars \citep{fan2013}, 
the size of the optical-NIR emission region $R$ is evaluated \citep{zhang2012} as 
$R<c\delta\Delta{t}/(1+z)=10\times0.38/(1+0.84)$~light-days$\sim2.1$~light-days$=5.4\times10^{15}$~cm, 
which corresponds to $1.2\times10^{3}$ times the Schwarzschild radius 
for a $1.5\times10^{7}$~M$_\odot$ BH ($R_s=4.4\times10^{12}$~cm) and 
is larger than the rest-frame UV-optical emitting region. 

%%%%%%%%%%%%%%%%%%%%%%%%%%%%%%%%%%%%%%%%
\subsection{Simultaneous Spectral Energy Distribution}\label{sec:sec_sed}
%%%%%%%%%%%%%%%%%%%%%%%%%%%%%%%%%%%%%%%%

The light curves sometimes show clear intranight variability 
by a factor of up to $\sim2$, 
so the photometry used to construct SEDs should be almost simultaneous. 
All the colors and SEDs shown below are measured for data taken 
almost simultaneously, i.e., with time intervals shorter than $2$~hours ($0.083$~days). 

\begin{figure*}[htbp!]
\begin{center}
	\includegraphics[angle=270,width=141mm]{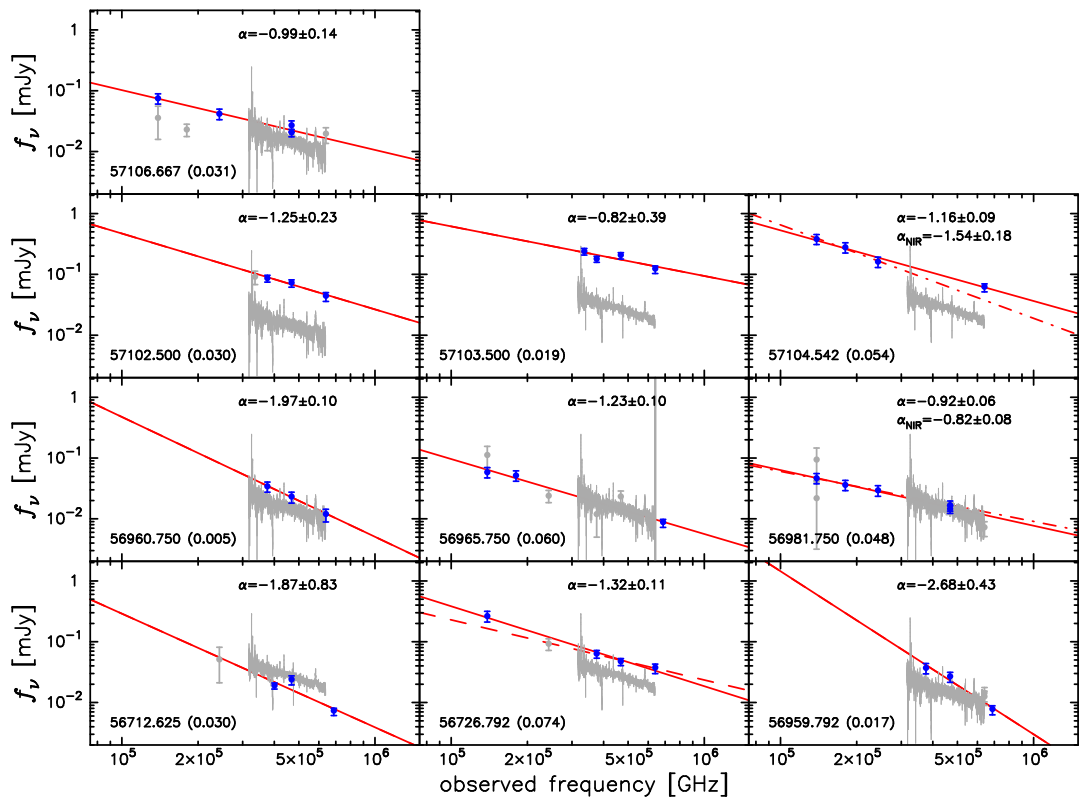}
\end{center}
\caption{
The SEDs from NIR to optical during the simultaneous observations in blue filled circles. 
Observing times in MJD are shown in bottom-left and the time baseline among the observing points 
are shown in the parenthesis in a unit of day. 
The fitted lines are shown in red. 
Note that gray filled circles are data points with low signal-to-noise ratios (S/N$<3$) and not used for the fitting. 
Solid, dashed, and dot-dashed lines indicate that the SEDs in NIR and optical, optical, and NIR wavelengths are fitted. 
The FOCAS spectra in the two phases ($\alpha_{\nu,{\rm{opt}}}=-1.40$, the brightest and faintest) are also shown in gray alternately 
for a reference purpose. 
The Galactic extinction are corrected for the data points and power-law indices calculated here.
\label{fig:fig_sed}}
\end{figure*}

\begin{figure*}[htbp!]
\begin{center}
	\includegraphics[angle=270,width=141mm]{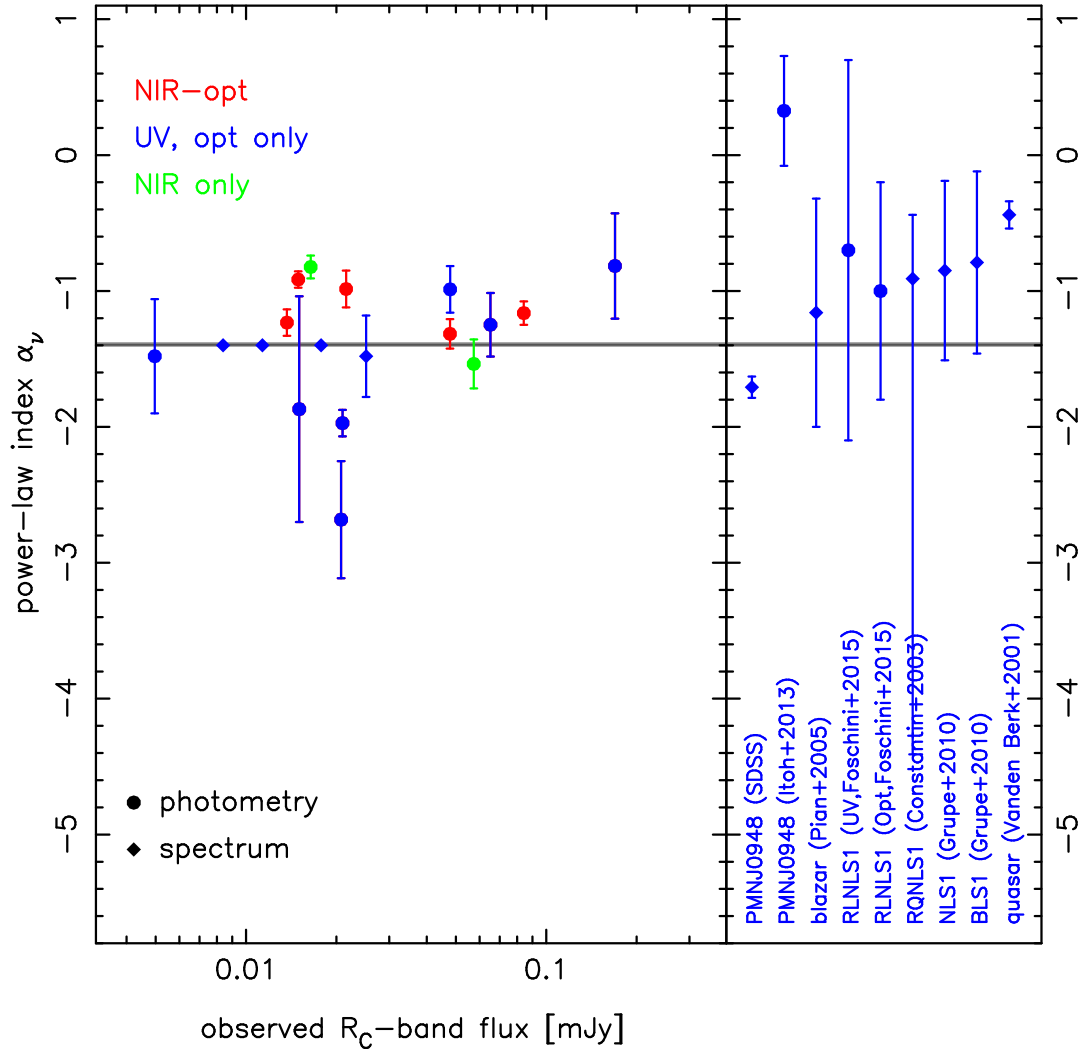}
\end{center}
\caption{
{\it (Left)}: 
Power-law index versus $R_C$-band flux for SDSS~J1100+4421. 
Filled circles are values obtained with the simultaneous imaging data. 
Different colors (red, blue, and green) indicate different wavelength ranges (optical and NIR, optical and NIR in the observed frame) 
fitted to calculated the power-law indices. 
Filled boxes indicate the power-law indices obtained from the FOCAS and GMOS spectra. 
The leftmost filled circle is the power-law index from the SDSS photometry. 
The averaged $\alpha_{\nu}$ is shown in gray shaded region. 
The Galactic extinction are corrected when we calculate the power-law indices.
{\it (Right)}: Power-law indices of various kinds of AGN from the literature. 
Indices of 
the optical SDSS spectrum of PMN~J0948+0022, 
optical imaging of PMN~J0948+0022 \citep{itoh2013}, 
UV spectra of blazars \citep{pian2005}, 
UV imaging of radio-loud NLS1s \citep{foschini2015}, 
optical spectra of radio-loud NLS1s \citep{foschini2015}, 
UV spectra of radio-quiet NLS1s \citep{constantin2003}, 
UV imaging of narrow-line and broad line Seyferts \citep{grupe2010}, and 
optical SDSS spectra of quasars \citep{vandenberk2001} are shown. 
The bars indicate 
the root-mean-square of $\alpha_{\nu}$ for \citet{grupe2010}, 
the systematic uncertainty of $\alpha_{\nu}$ for quasars \citep{vandenberk2001}, 
and the ranges of the $\alpha_{\nu}$ values for the rest of the papers. 
\label{fig:fig_sed_pl}}
\end{figure*}

We made simultaneous SEDs at 10 epochs 
as shown in Figure~\ref{fig:fig_sed},
including instances during all the three observing epochs.
We fit each SED with a single power-law ($f_\nu\propto\nu^{\alpha_\nu}$). 
The power-law indices $\alpha_\nu$ are plotted 
as a function of observed or interpolated $R_C$-band brightness 
in Figure~\ref{fig:fig_sed_pl}. 
In addition to fitting the photometric SEDs, the FOCAS and GMOS optical spectra were also fitted with a single power-law. 
We note that the fitting results for the FOCAS spectra ($\alpha_\nu=-1.40\pm0.01$) 
were presented in \citet{tanaka2014}. 
The obtained power-law index of the GMOS spectrum is $\alpha_\nu=-1.48\pm0.10$ but 
the spectrum is not corrected for the differential slit loss due to atmospheric differential refraction. 
The possible differential slit loss is about 20\%, causing an overestimate of $\alpha_\nu$. 
We here do not estimate the exact factor and set the error of $\alpha_\nu$ from the GMOS spectrum to be $0.3$, 
roughly corresponding to the 20\% difference. 
(i.e., $\alpha_\nu=-1.48\pm0.30$ instead of $\alpha_\nu=-1.48\pm0.10$). 
These fitting results for the spectra are included in Figure~\ref{fig:fig_sed_pl}. 

Overall, the obtained power-law indices are consistent with being identical,
although some of the indices are somewhat steep (small $\alpha_\nu$). 
This indicates that the origin of the optical-NIR 
emission (or spectrum) is the same regardless of brightness and observing epochs. 
The weighted average of the indices among our own monitoring data is $\alpha_\nu=-1.39\pm0.02$ as shown in gray shade in Figure~\ref{fig:fig_sed_pl}. 

%%%%%%%%%%%%%%%%%%%%%%%%%%%%%%%%%%%%%%%%
%\clearpage
%%%%%%%%%%%%%%%%%%%%%%%%%%%%%%%%%%%%%%%%

%%%%%%%%%%%%%%%%%%%%%%%%%%%%%%%%%%%%%%%%
%%%%%%%%%%%%%%%%%%%%%%%%%%%%%%%%%%%%%%%%

\begin{figure*}[htbp!]
\begin{center}
	\includegraphics[angle=270,width=151mm]{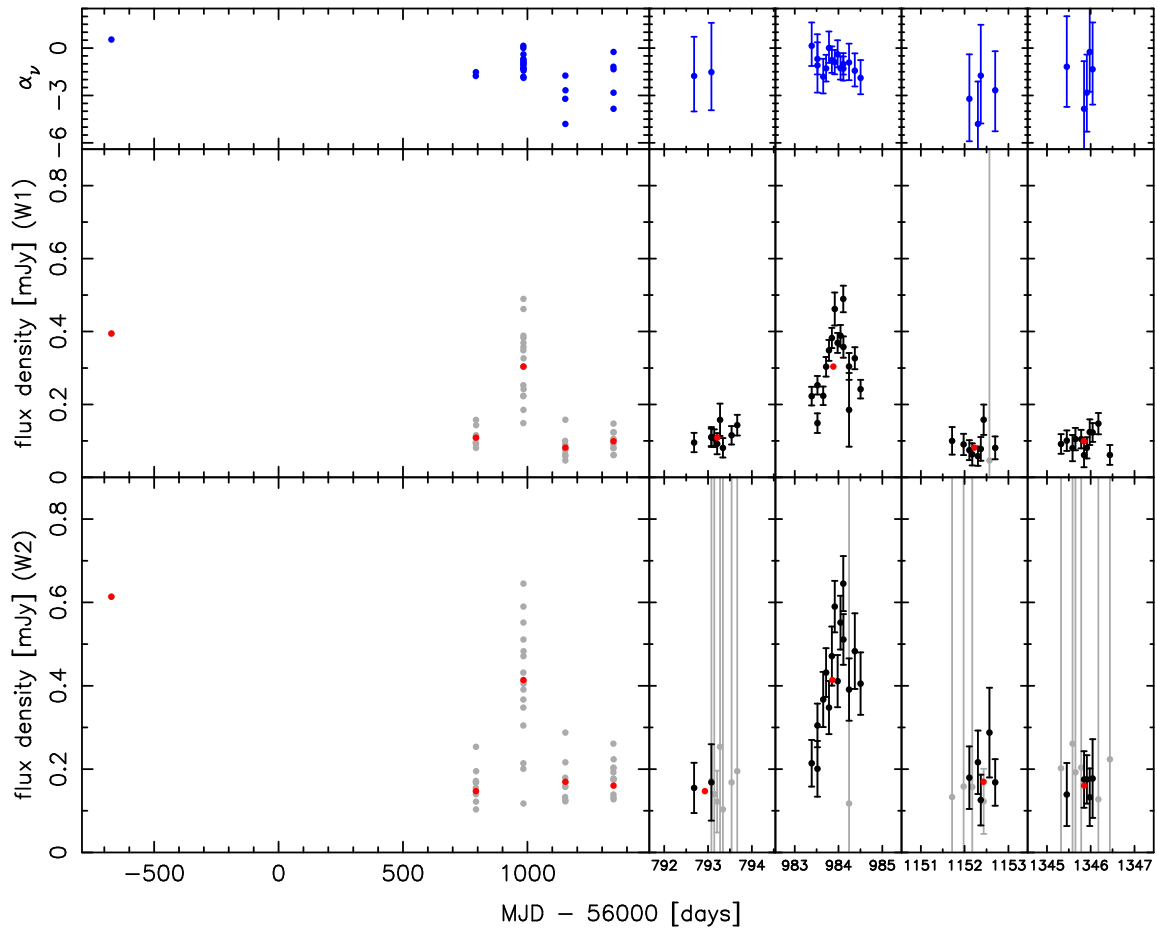}
\end{center}
\caption{
{\it (Top)}: Power-law indices calculated with the WISE $3.4~\mu$m and $4.6~\mu$m-band data as a function of time. 
{\it (Middle), (Bottom)}: 
WISE light curves of J1100+4421 in the $3.4~\mu$m (W1) and $4.6~\mu$m (W2) -bands are shown in gray. 
The left panels are for the whole range of the WISE data while 
the other panels are magnified views of the four observing epochs of the NEOWISE Reactivation mission.  
In the four right panels, data points with $<0.5$~mag errors are shown in black.
Weighted averages of flux during each epoch is shown in red. 
The error bars are shown only in the magnified panels. 
\label{fig:fig_wiselc}}
\end{figure*}

In addition to our data, 
archival mid-infrared (MIR) data obtained with Wide-Field Infrared Survey Explorer (WISE) in the $3.4~\mu$m (W1) and $4.6~\mu$m-bands (W2) are available. 
These data were taken during the NEOWISE Reactivation mission \citep{mainzer2014}. 
We made WISE light curves of J1100+4421 and nearby reference objects 
from the {\it Single Exposure (L1b) Source Table}. 
The WISE magnitudes in the Vega system obtained from the catalogs are converted to AB magnitudes using 
the equations in \citet{jarrett2011}, 
$W1_{\rm{AB}}=W1_{\rm{Vega}}+2.699$ and 
$W2_{\rm{AB}}=W2_{\rm{Vega}}+3.339$. 
By combining this with the WISE data shown in \citet{tanaka2014}, 
J1100+4421 shows clear variability around 
56984 (Nov. 23, 2014)
both in the W1 and W2 bands (Figure~\ref{fig:fig_wiselc}),
while the reference objects are stable in flux within the error bars. 
The power-law indices derived from the two-band WISE data 
are $\alpha_{\nu,\rm{MIR}}\sim-1.4$ at all epochs. 
On days of MJD$=56983-4$, the optical-MIR power-law index are also $\alpha_{\nu,\rm{opt,MIR}}\sim-1.4$.

%%%%%%%%%%%%%%%%%%%%%%%%%%%%%%%%%%%%%%%%
%%%%%%%%%%%%%%%%%%%%%%%%%%%%%%%%%%%%%%%%

%%%%%%%%%%%%%%%%%%%%%%%%%%%%%%%%%%%%%%%%
\subsection{Host Galaxy and Environment}\label{sec:sec_host_result}
%%%%%%%%%%%%%%%%%%%%%%%%%%%%%%%%%%%%%%%%

\begin{figure*}[!htbp]
\begin{center}
	\includegraphics[angle=0,width=83mm]{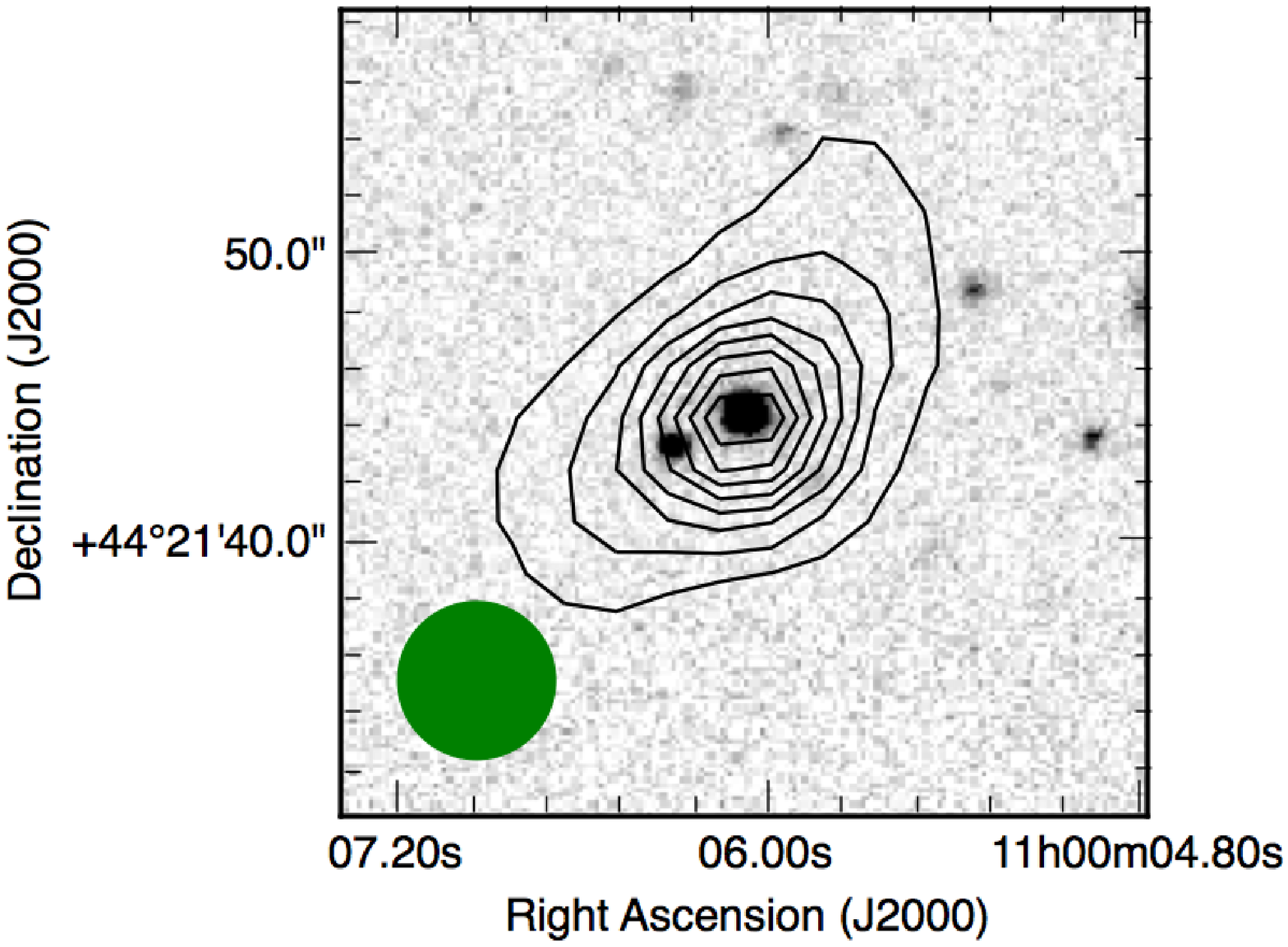}
	\includegraphics[angle=0,width=83mm]{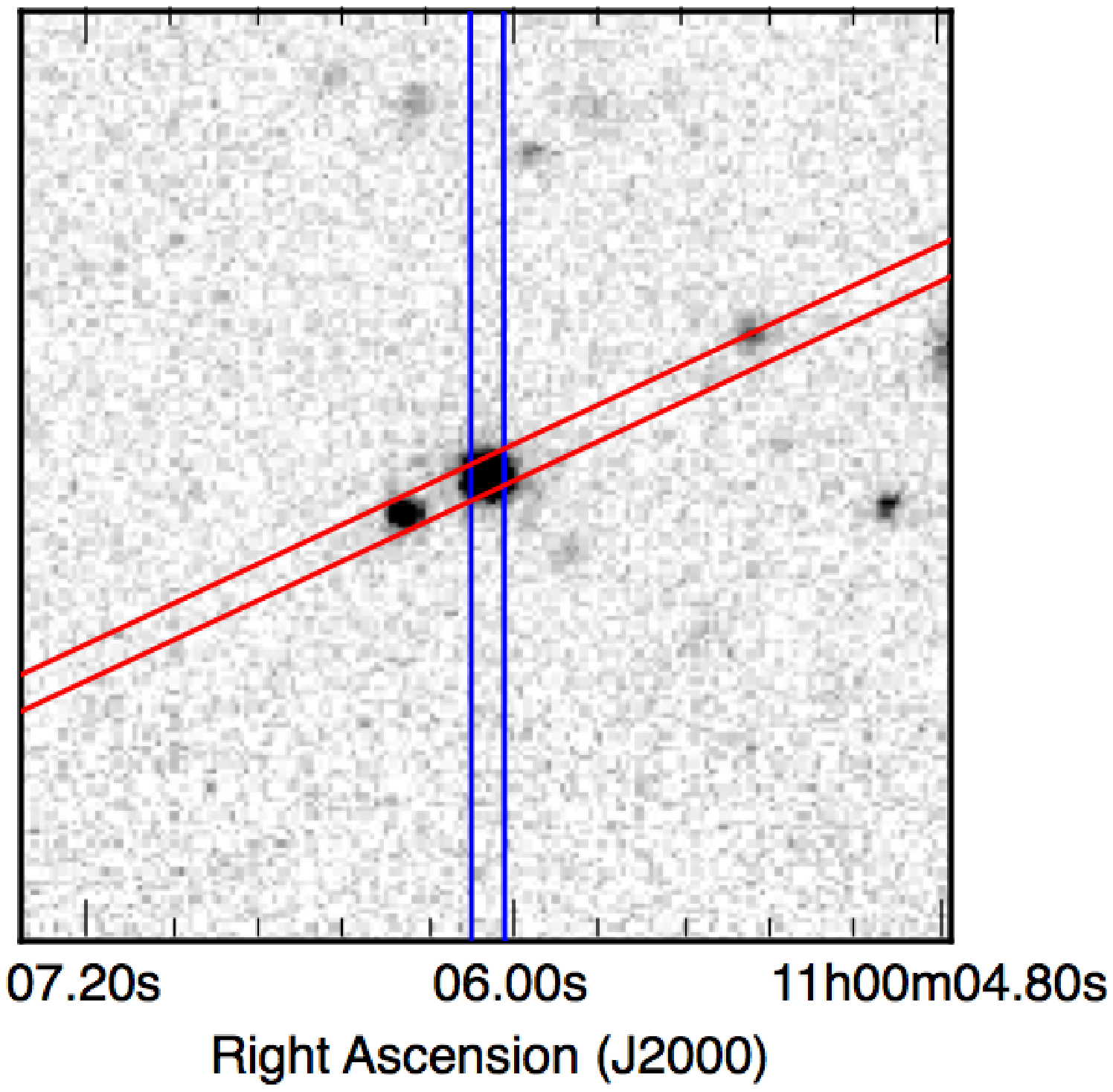}
\end{center}
\caption{
{\it (Left)}: 
Subaru/HSC $z$-band image in gray scale overlaid with the FIRST radio data in contour. 
The central object in this figure is J1100+4421. 
Beam size of the FIRST image (5.4~arcsec) is shown at the left-bottom part of the figure in green. 
{\it (Right)}: 
The slits of the FOCAS and GMOS observations are shown in blue and red, respectively. 
\label{fig:fig_hscz_radio}}
\end{figure*}

Host galaxies of blazars are known to be massive elliptical galaxies 
(\cite{sbarufatti2005}; \cite{falomo2014}) 
while those of radio-loud NLS1 galaxies are star-forming galaxies (\cite{zhou2007}; \cite{anton2008}; 
\cite{tavares2014} \cite{caccianiga2015}).
The central BH mass of J1100+4421 is measured using the H$\beta$ and Mg$_{\rm{II}}$ emission lines 
and found to be as small as 
$1.0-1.5\times10^{7}$~M$_\odot$ \citep{tanaka2014}. 
Considering this small BH mass, 
the host galaxy properties are expected to be similar to those of RLNLS1, rather than those of blazars. 
In addition, a large spatial extension of radio emission from this host galaxy is detected in FIRST data \citep{becker1995}. 
We investigate here the host galaxy properties using the Subaru HSC images 
with sub-arcsec spatial resolution (see Figure~\ref{fig:fig_hscz_radio}). 

%%%%%%%%%%%%%%%%%%%%%%%%%%%%%%%%%%%%%%%%
%%%%%%%%%%%%%%%%%%%%%%%%%%%%%%%%%%%%%%%%

The seeing sizes (FWHM) 
of the Subaru HSC images
are 
0.56~arcsec in $g$-band and 
0.51~arcsec in $z$-band. 
The FWHM sizes of the object are 
0.58~arcsec in $g$-band and 
0.55~arcsec in $z$-band. 
These are almost unresolved and dominated by the nucleus.
We note that the HSC magnitudes are 
$g=21.71\pm0.12$~mag and $z=20.57\pm0.11$~mag, respectively, 
both of are $\sim0.4$~mag brighter than the SDSS magnitudes, which have photometric uncertainties of $\Delta{g}=0.11$ and $\Delta{z}=0.25$. 
We also note that these two HSC observations were done at different epochs. 

Compared with the extended ($\sim10$~arcsec; \cite{gabanyi2017}) radio emission detected in the FIRST data, 
the optical sizes measured above are much smaller. 
However, the optical sizes are measured using broad-band imaging data,
 and most of the light 
is continuum emission, including components from the AGN and the host galaxy.
In the $z$-band bandpass, [O$_{\rm{~III}}$] emission lines are included as well as the H$\beta$ emission line,
and we here set a brightness constraint on the extended [O$_{\rm{~III}}$] emission. 
As seen in the HSC $z$-band image, no extended structure aligned to the radio structure is detected. 
Considering the width of the bandpass, we convert 
the upper limit of $z$-band surface brightness $z_{\rm{lim}}=29.4~$ [mag arcsec$^{-2}$] ($3\sigma$) 
to the upper limit of [O$_{\rm{~III}}$] surface brightness $m(\rm{[O_{\rm{~III}}]})_{\rm{lim}}=24.4~$ [mag arcsec$^{-2}$] ($3\sigma$) 
by assuming that the [O$_{\rm{~III}}$] spectral widths are the same ($\sim10$\AA\ FWHM) as that of the central source spectrum. 
This limit corresponds to a flux density of 
$2.1\times10^{-18}$~[erg s$^{-1}$ cm$^{-2}$ arcsec$^{-2}$] 
and is smaller than the [O$_{\rm{~III}}$] flux detected for the central source by a factor of $\sim600$. 
In addition to the imaging data, we examine the spatial extension of the [O$_{\rm{~III}}$] emission lines 
along the slits 
detected in our spectroscopic data with FOCAS and GMOS. 
The sizes perpendicular to the slits are almost the same as or slightly extended than those of the continua, 
also indicating that the [O$_{\rm{~III}}$] emitting region size is not as large as that in radio wavelength. 
We also note that there are no clear blue-shifted [O$_{\rm{~III}}$] emission lines 
in any of our FOCAS and GMOS spectra.

%%%%%%%%%%%%%%%%%%%%%%
%%%%%%%%%%%%%%%%%%%%%%
%%%%%%%%%%%%%%%%%%%%%%
\section{Discussion}\label{sec:sec_discussions}
%%%%%%%%%%%%%%%%%%%%%%
%%%%%%%%%%%%%%%%%%%%%%
%%%%%%%%%%%%%%%%%%%%%%

%%%%%%%%%%%%%%%%%%%%%%
\subsection{Contributions from Host Galaxy and Accretion Disk}\label{sec:sec_originemission}
%%%%%%%%%%%%%%%%%%%%%%%%%%%%%%%%%%%%%%%%%%

We first evaluate the luminosity contribution of 
the host galaxy and the accretion disk. 
We note that the observed wavelength range corresponds to $0.2$~$\mu$m to $1.2$~$\mu$m in the rest-frame, 
UV to short NIR wavelengths. 

The spectra do not show 
any significant absorption lines which could indicate host galaxy components. 
The faintest photometry among all of our data are the 5-band SDSS values. 
Although the measurement errors are not small ($\Delta{m}=0.11-0.40$~mag), 
the measurements in the five different bands are almost simultaneous, 
the observation times ranging from MJD$=52722.1509$ to MJD$=52722.1542$ (48~minutes difference on March 23, 2004), and 
the fitted power-law index is 
$\alpha_{\nu}=-1.48\pm0.42$ (the leftmost point in the left panel of Figure~\ref{fig:fig_sed_pl});
this is consistent with the power-law indices computed at other epochs. 
Although there is a possibility that the host galaxy color is similar to that of the varying AGN component, 
the contribution from the host galaxy, especially during the bright phases when we detected it with 1-2m class telescopes,
 is expected to be small. 
This is consistent with the fact that host galaxies (bulge components) of low-$z$ NLS1s are as faint as $M_{B,\rm{bulge}}=-18.5$~mag \citep{botte2004}, 
corresponding to $m_{i}\sim25.2$~mag at this redshift, 
much fainter than the detection limits of most of our data. 

The second item to be examined is the luminosity of the accretion disk. 
The BH mass is estimated to be $1.0-1.5\times10^{7}$~M$_\odot$, 
based on the line luminosity and widths of the Mg$_{\rm{II}}$ and H$\beta$ emission lines 
~\citep{tanaka2014}. 
The bolometric luminosity can be explained by sub-Eddington accretion ($\sim30$\%). 
For a BH with this small mass and normal (or somewhat high) Eddington ratio, 
the peak wavelength of the accretion disk semi-blackbody emission is 
UV to soft X-ray 
if we assume that the object has a standard accretion disk \citep{shakura1973}. 
This means that the luminosity of the UV-optical wavelength region is expected to be $f_\nu\propto\nu^{1/3}$ ($\alpha_\nu=1/3$) 
which is inconsistent with, much harder than, the obtained power-law index $\alpha_\nu\sim-1.4$. 
This indicates the accretion disk component does not significantly contribute to the observed flux 
in the observed wavelength range. 

%%%%%%%%%%%%%%%%%%%%%%
\subsection{Synchrotron Radiation in optical-NIR Wavelength Regions}\label{sec:sec_synchrotron}
%%%%%%%%%%%%%%%%%%%%%%

Considering the small luminosity contribution from the host galaxy and accretion disk, 
a plausible explanation for the varying luminosity with constant power-law indices
would be synchrotron emission by high-energy electrons inside a relativistic jet. 

Following the conventional synchrotron theory (see \cite{rybicki1979} for details), 
we can derive the power-law index of electron distribution ($dN_{e}/dE\propto E^{-p}$) 
from the observed optical-IR spectrum as $p=-2\alpha+1=3.8$. 
This is much softer than the prediction of standard diffusive shock acceleration theory, 
which predicts the power-law index of $p=2.0$ (e.g, \cite{blandford1978}). 
We note that such a soft electron distribution is easily achieved by e.g., radiative cooling, 
inefficient acceleration and so on, and hence 
the electron power-law index of $\sim3.8$ estimated here is not unusual. 
Indeed, it is well known that similar soft electron distribution is 
derived in many blazars by applying synchrotron plus inverse-Compton 
modeling of the broad-band spectral energy distribution from radio to $\gamma$-ray (see e.g, \cite{ghisellini2010}). 

The emission mechanism can be investigated with higher-energy observational data where 
the synchrotron self-Compton (SSC) radiation dominates. 
We checked Fermi Large Area Telescope (LAT) source catalog (3FGL; Fermi LAT 4-Year Point Source Catalog; \cite{acero2015}) 
and found that no object is detected at the position of J1100+4421;
and no source appears in the 2FGL catalog investigated in \citet{tanaka2014}. 
The Fermi/LAT upper limit in $\nu F_{\nu}$ units is a few times $10^{-12}$~erg~s$^{-1}$~cm$^{-2}$ \citep{acero2015}. 
This upper limit 
is roughly comparable to the expected SSC emission and can be explained by the synchrotron/SSC model \citep{chiang2002} if we assume $p=3.8$. 

The observed optical-NIR SEDs are well described by synchrotron emission, which 
indicates that the peak frequency of the synchrotron emission is 
lower than the observed wavelength range, $\nu_{\rm{peak}}<10^{15}$~Hz,  
and so the object belongs to a low-frequency-peaked blazar population. 
The observed optical-NIR colors of J1100+4421 at $z=0.840$,
obtained almost simultaneously, are 
$g-J\sim1.1$, 
$R-J\sim0.8$, and 
$I-J\sim0.6$ in the AB system, 
roughly corresponding to 
$g-J\sim2.1$, 
$R-J\sim1.5$, and 
$I-J\sim1.1$ in the Vega system. 
These colors are at the blue end of the distribution for blazars at $z\sim0.02-1.8$ 
shown in \citet{ikejiri2011} 
and consistent with both the colors of higher- or lower-frequency-peaked populations. 
The large variability amplitude 
    ($\Delta{g}\sim3.5$~mag, a factor of $\sim30$, between the SDSS and our March 2015 photometry) 
also might indicate that J1100+4421 is similar to low $\nu_{\rm{peak}}$ blazars 
because there are no high $\nu_{\rm{peak}}$ blazars showing such a large amplitude of variability \citep{ikejiri2011}. 
The small color variability within our observations, i.e. almost constant power-law indices, 
is consistent with the object belonging to a low $\nu_{\rm{peak}}$ blazar population. 

A comparison with the spectral indices of other AGN is shown in the right panel of Figure~\ref{fig:fig_sed_pl}. 
The power-law index $\alpha_\nu$ of J1100+4421 in the rest-frame UV to short NIR wavelengths 
 is consistent with 
those in the observed UV wavelengths for 
blazars at $z=0.15-1.41$ ($\alpha_{\nu,{\rm{UV}}}=-2.00$--$-0.32$; \cite{pian2005}), 
RLNLS1 at $z=0.06-0.80$ ($\alpha_{\nu,{\rm{UV}}}=-0.7\pm1.4$, median $-0.7$; \cite{foschini2015}), 
RLNLS1 at $z=0.06-0.80$ ($\alpha_{\nu,{\rm{opt}}}=--1.0\pm0.8$, median $-0.8$; \cite{foschini2015}), 
RQNLS1 at $z=0.01-0.26$ ($\alpha_{\nu,{\rm{UV}}}=-4.41$--$-0.44$, median $-0.91$; \cite{constantin2003}), 
broad-line Seyferts at $<z>=0.086$ ($\alpha_{\nu,{\rm{UV}}}=-0.79\pm0.67$, median $-0.61$; \cite{grupe2010}), and 
narrow-line Seyferts at $<z>=0.087$ ($\alpha_{\nu,{\rm{UV}}}=-0.85\pm0.66$, median $-0.65$; \cite{grupe2010}) 
We note that most of these values are the range of the $\alpha_{\nu}$ values, not the root-mean-square, except for the values 
in \citet{grupe2010}. 
On the other hand, the power-law index of J1100+4421 is clearly different from those of SDSS quasars 
($\alpha_{\nu}=-0.44$, $\sim0.1$ uncertainty) shown in \citet{vandenberk2001} 
and most ($>90$\%) of the SDSS quasars have harder UV-optical spectra than J1100+4421 \citep{shen2011}. 
This difference also supports our hypothesis that the emission of J1100+4421 in optical-NIR wavelength regions 
arises from synchrotron radiation in the relativistic jet, not an accretion disk. 

The famous Fermi LAT-detected NLS1, PMN~J0948+0022 \citep{abdo2009}, 
a low $\nu_{\rm{peak}}$ radio/$\gamma$-loud NLS1 at $z=0.585$, shares observational properties similar to J1100+4421 
including kpc-scale radio emission \citep{doi2012},
but lacks the high [O$_{\rm{~III}}$]/H$\beta$ flux ratio of J1100+4421. 
If J1100+4421 follows the general distributions observed for low $\nu_{\rm{peak}}$ blazars in \citet{ikejiri2011}, 
a high degree of polarization might be expected for J1100+4421 
as was indeed detected from PMN~J0948+0022 \citep{itoh2013}.

PMN~J0948+0022 shows a different behavior in 
its power-law indices than J1100+4421. 
The extinction-corrected ($A_V=0.263$~mag; \cite{schlegel1998}; assuming the extinction law given by \cite{cardelli1989}) 
SDSS spectrum of PMN~J0948+0022 gives a power-law index $\alpha_{\nu}=-1.77\pm0.07$. 
On the other hand, 
\citet{itoh2013} measured an optical power-law index of $0<\alpha_{\nu}<1$ during a brighter phase of PMN~J0948+0022 
after the detection of a $\gamma$-ray flare with Fermi LAT. 
They indicated that the index varies at different brightness phases, 
which is a well known behavior known as the ``bluer-when-brighter'' trend for blazars (e.g., \cite{ikejiri2011}) and quasars (e.g., \cite{kokubo2014});
we note that J1100+4421 does not show such a change in its power-law indices. 

%%%%%%%%%%%%%%%%%%%%%%%
\subsection{Host Galaxy and Environment}\label{sec:sec_hostgal}
%%%%%%%%%%%%%%%%%%%%%%%

As pointed out in previous papers, 
there is a hypothesis that 
radio-loud NLS1s, which are an AGN population 
%as a group 
mostly similar to J1100+4421, are 
products of interaction 
(\cite{zhou2007}; \cite{anton2008}). 
Therefore, we consider here whether or not this hypothesis is valid for J1100+4421. 
For more massive systems like quasars, the existence of extended narrow emission line regions sometimes indicates
recent minor merger events and AGN triggering (\cite{fu2009}; \cite{matsuoka2012}). 
An observational study on extended narrow-line regions in nearby Seyfert galaxies \citep{keel2012} 
also shows that a significant fraction of the sample galaxies are interacting or merging systems. 
However, as shown in Section \ref{sec:sec_host_result}, 
we tentatively conclude that an extended [O$_{\rm{~III}}$] emitting region is {\it not} present in our optical broad-band imaging data. 

This indicates the small ($<1$~arcsec FWHM corresponding to several~kpc) size of the [O$_{\rm{~III}}$] emitting regions. 
This small size of the [O$_{\rm{~III}}$] narrow-line region itself is partly consistent with 
the notion that large, extended narrow-line regions are usually associated with 
AGN which have low Eddington ratios and/or high BH masses \citep{matsuoka2012}. 

The power-law index of this object in the radio frequency is hard to measure 
because of non-simultaneous observations as shown in Table~1 of \citet{tanaka2014}. 
The indices naively measured (i.e., ignoring the time differences between the observations) 
indicate both possibilities that J1100+4421 belongs to flat- or steep-spectrum population. 
If J1100+4421 is a steep-spectrum radio-loud AGN which is indicated by the largely extended radio lobe, 
the radio-loud nature, which is one of the unique points of J1100+4421, 
may be inconsistent with the fact that 
a high fraction (about 60\%) of 
steep-spectrum radio-loud quasars 
may host extended narrow-line regions,
although not all of them do \citep{matsuoka2012}. 

We identified a galaxy 
at (RA, Dec) = (11:00:06.30, +44:21:43.4), to the east of J1100+4421 
by 2.7~arcsec (21~kpc at this redshift, $z=0.840$) in our deep HSC imaging data. 
The magnitudes of this galaxy are $g=25.32\pm0.36$~mag and $z=22.15\pm0.09$~mag 
in Novermber 2014, 
a factor of %several 
18.4 and 3.0 
fainter than the faintest phases of J1100+4421, respectively. 
We took a GMOS optical spectrum of this galaxy with a 1.0~arcsec-width slit along J1100+4421 as described in Section~\ref{sec:sec_followup_optspec}. 
The spectrum does not show any significant continuum because of the faintness of the galaxy and the short exposure, and 
does not show any emission lines at the same redshift as that of J1100+4421. 
Hence it is unclear whether or not this galaxy interacts with J1100+4421. 
As pointed by \citet{doi2012}, 
RLNLS1s with extended radio structure, with which J1100+4421 shares similar properties,
are thought to be at the final stage of evolution of NLS1s, 
possibly as they evolve 
to broad-line, more massive, AGN. 
With this hypothesis in mind, the possible non-detection of extended narrow-line regions might be consistent with 
preferred detections in massive systems. 

%%%%%%%%%%%%%%%%%%%%%%
%%%%%%%%%%%%%%%%%%%%%%
%%%%%%%%%%%%%%%%%%%%%%
\section{Summary}\label{sec:sec_summary}
%%%%%%%%%%%%%%%%%%%%%%
%%%%%%%%%%%%%%%%%%%%%%
%%%%%%%%%%%%%%%%%%%%%%

We carried out monitoring campaign observations 
at optical and NIR wavelengths 
for the radio-loud AGN J1100+4421 
with a BH mass of $1.0-1.5\times10^{7}$~M$_{\odot}$. 
The light curves and simultaneous SEDs are shown and 
rapid variability behaviors are detected in each of the three observing campaigns. 
The object shows large variability during our observations, changing by a factor of $\sim30$ while 
its power-law indices remain unchanged ($\alpha_{\nu}\sim-1.4$ where $f_{\nu}\propto\nu^{\alpha_{\nu}}$). 
All of the observational results are consistent with the hypothesis that the optical-NIR emission originates from 
synchrotron radiation in a relativistic jet. 
The small SED changes indicate that J1100+4421 belongs to a low $\nu_{\rm{peak}}$ blazar population. 
The marginal spatial resolution in sub-arcsec optical imaging data taken with Subaru HSC,
showing no large, extended [O$_{\rm{~III}}$] narrow-line region, 
is smaller than the extended radio emission detected in the FIRST survey data. 
A newly obtained optical spectrum of a possible companion galaxy does not support 
   the idea that the galaxy is at the same redshift of J1100+4421, or the merging hypothesis 
for triggering the AGN-related activity of J1100+4421. 
To understand the nature of J1100+4421, which does not share all its properties with any other set of objects like NLS1s, 
further observations including 
multi-wavelength, high-resolution radio VLBI \citep{gabanyi2017}, 
deep narrow-line region imaging or integral-field spectroscopy, 
more densely-sampled monitoring, 
and polarization are required. 

%%%%%%%%%%%%%%%%%%%%%%%%%%%%%%%%%%%%%%%%%%%%
%%%%%%%%%%%%%%%%%%%%%%%%%%%%%%%%%%%%%%%%%%%%
%%%%%%%%%%%%%%%%%%%%%%%%%%%%%%%%%%%%%%%%%%%%

%\acknowledgements

%%%%%%%%%%%%%%%
%%%%%% LC %%%%%
%%%%%%%%%%%%%%%
\begin{longtable}{cccccc}
  \caption{Summary of Optical and Near-Infrared Photometric Measurements of J1100+4421. Galactic extinctions are not corrected. 
\label{tab:tab_obs_niropt}
  }
  \hline
      telescope & instrument & filter & MJD & date (UT) & magnitude \\%& $\Delta{t}$[days],$\Delta{m}$[mag],flux-ratio,$\tau$\\
	\hline
\endfirsthead
  \hline
      telescope & instrument & filter & MJD & date (UT) & magnitude\\
	\hline
\endhead
  \hline
\endfoot
  \hline
\endlastfoot
  \hline
	\input{kiss14k_lc_ver8_mjdsort4}
\hline
\end{longtable}

%%%
\begin{ack}
This work was supported by the Optical and Near-infrared Astronomy Inter-University Cooperation Program
and the Grants-in-Aid of the 
Ministry of Education, Science, Culture, and Sport 
23740143, 25800103, 16H02158 
15H02075, 15H00788, 
24740117, 25103515, 23740157, 
15J10324, 19047003, 25707007.
This work was partially carried out by the joint research program of the Institute for Cosmic Ray Research (ICRR), University of Tokyo,
based in part on data collected at Subaru Telescope, which is operated by the National Astronomical Observatory of Japan and  
data obtained at the Gemini Observatory via the time exchange program between Gemini and the Subaru Telescope, 
and 
also based on data obtained by SDSS-III. 
The Gemini Observatory is operated by 
the Association of Universities for Research in Astronomy, Inc., under a cooperative agreement with the NSF on behalf of the Gemini partnership: the National Science Foundation (United States), the National Research Council (Canada), CONICYT (Chile), Ministerio de Ciencia, Tecnolog\'{i}a e Innovaci\'{o}n Productiva (Argentina), and Minist\'{e}rio da Ci\^{e}ncia, Tecnologia e Inova\c{c}\~{a}o (Brazil).
SDSS-III is managed by the Astrophysical Research Consortium for the Participating Institutions of the SDSS-III Collaboration including the University of Arizona, the Brazilian Participation Group, Brookhaven National Laboratory, Carnegie Mellon University, University of Florida, the French Participation Group, the German Participation Group, Harvard University, the Instituto de Astrofisica de Canarias, the Michigan State/Notre Dame/JINA Participation Group, Johns Hopkins University, Lawrence Berkeley National Laboratory, Max Planck Institute for Astrophysics, Max Planck Institute for Extraterrestrial Physics, New Mexico State University, New York University, Ohio State University, Pennsylvania State University, University of Portsmouth, Princeton University, the Spanish Participation Group, University of Tokyo, University of Utah, Vanderbilt University, University of Virginia, University of Washington, and Yale University. 
This publication makes use of data products from the Two Micron All Sky Survey, which is a joint project of the University of Massachusetts and the Infrared Processing and Analysis Center/California Institute of Technology, funded by the National Aeronautics and Space Administration and the National Science Foundation.
This paper makes use of software developed for the Large Synoptic Survey Telescope. We thank the LSST Project for making their code available as free software at http://dm.lsstcorp.org. Funding for SDSS-III has been provided by the Alfred P. Sloan Foundation, the Participating Institutions, the National Science Foundation, and the U.S. Department of Energy Office of Science. The SDSS-III web site is http://www.sdss3.org/. 

We also appreciate a kind help by Prof.~Michael W. Richmond for improving the English grammar of the manuscript 
and a private communication with Dr.~Luigi Foschini. 

\end{ack}

\bibliography{Morokuma2017KISS14k_arxiv201707a}

\end{document}

%% file: kiss14k_lc_ver8_mjdsort4.tex
Kiso & KWFC & $ u $ & 56958.80 & 2014-10-28 & $> 20.27$     \\
Kiso & KWFC & $ u $ & 56964.82 & 2014-11-03 & $> 20.41$     \\
Kiso & KWFC & $ u $ & 56968.84 & 2014-11-07 & $> 18.86$     \\
Kiso & KWFC & $ u $ & 56976.74 & 2014-11-15 & $> 19.56$     \\
Kiso & KWFC & $ u $ & 56977.82 & 2014-11-16 & $> 20.36$     \\
\hline
Subaru & FOCAS & $ B $ & 56712.27 & 2014-02-24 & $21.17\pm0.10$      \\
Subaru & FOCAS & $ B $ & 56712.55 & 2014-02-24 & $21.72\pm0.10$      \\
Subaru & FOCAS & $ B $ & 56712.63 & 2014-02-24 & $21.78\pm0.10$      \\
Pirka & MSI & $ B $ & 56954.82 & 2014-10-24 & $> 20.03$     \\
Pirka & MSI & $ B $ & 56955.82 & 2014-10-25 & $> 16.95$     \\
Pirka & MSI & $ B $ & 56959.80 & 2014-10-29 & $21.71\pm0.14$      \\
Pirka & MSI & $ B $ & 56965.78 & 2014-11-04 & $21.58\pm0.12$      \\
Pirka & MSI & $ B $ & 56983.84 & 2014-11-22 & $21.16\pm0.14$      \\
Pirka & MSI & $ B $ & 56984.83 & 2014-11-23 & $21.72\pm0.16$      \\
Pirka & MSI & $ B $ & 56986.84 & 2014-11-25 & $22.13\pm0.21$      \\
Pirka & MSI & $ B $ & 56987.83 & 2014-11-26 & $22.14\pm0.15$      \\
Pirka & MSI & $ B $ & 56988.79 & 2014-11-27 & $21.73\pm0.16$      \\
Kottamia & Kottamia & $ B $ & 57063.03 & 2015-02-10 & $> 19.93$     \\
Pirka & MSI & $ B $ & 57174.58 & 2015-06-01 & $> 19.25$     \\
Pirka & MSI & $ B $ & 57175.61 & 2015-06-02 & $> 16.22$     \\
\hline
Kiso & KWFC & $ g $ & 56325.66 & 2013-02-02 & $> 21.33$     \\
Kiso & KWFC & $ g $ & 56327.70 & 2013-02-04 & $> 21.06$     \\
Kiso & KWFC & $ g $ & 56347.59 & 2013-02-24 & $> 18.67$     \\
Kiso & KWFC & $ g $ & 56709.50 & 2014-02-21 & $> 20.90$     \\
Kiso & KWFC & $ g $ & 56710.51 & 2014-02-22 & $> 21.30$     \\
Kiso & KWFC & $ g $ & 56711.46 & 2014-02-23 & $19.73\pm0.16$      \\
Kiso & KWFC & $ g $ & 56711.51 & 2014-02-23 & $20.18\pm0.13$      \\
Kiso & KWFC & $ g $ & 56711.51 & 2014-02-23 & $20.22\pm0.20$      \\
Kiso & KWFC & $ g $ & 56711.55 & 2014-02-23 & $19.96\pm0.15$      \\
Kiso & KWFC & $ g $ & 56711.60 & 2014-02-23 & $19.82\pm0.14$      \\
Kiso & KWFC & $ g $ & 56712.72 & 2014-02-24 & $21.08\pm0.14$      \\
Kiso & KWFC & $ g $ & 56712.72 & 2014-02-24 & $21.19\pm0.14$      \\
Kiso & KWFC & $ g $ & 56713.63 & 2014-02-25 & $21.50\pm0.14$      \\
Kiso & KWFC & $ g $ & 56713.63 & 2014-02-25 & $21.48\pm0.14$      \\
Kiso & KWFC & $ g $ & 56714.46 & 2014-02-26 & $> 19.05$     \\
Kiso & KWFC & $ g $ & 56718.70 & 2014-03-02 & $> 20.34$     \\
Kiso & KWFC & $ g $ & 56726.82 & 2014-03-10 & $20.01\pm0.15$      \\
Kiso & KWFC & $ g $ & 56727.57 & 2014-03-11 & $> 20.62$     \\
Kiso & KWFC & $ g $ & 56728.49 & 2014-03-12 & $> 18.48$     \\
Kiso & KWFC & $ g $ & 56731.66 & 2014-03-15 & $> 20.23$     \\
Kiso & KWFC & $ g $ & 56732.77 & 2014-03-16 & $> 19.84$     \\
Kiso & KWFC & $ g $ & 56733.52 & 2014-03-17 & $> 20.04$     \\
Kiso & KWFC & $ g $ & 56738.74 & 2014-03-22 & $> 20.68$     \\
Kiso & KWFC & $ g $ & 56739.74 & 2014-03-23 & $> 20.86$     \\
Kiso & KWFC & $ g $ & 56744.51 & 2014-03-28 & $> 20.85$     \\
Kiso & KWFC & $ g $ & 56745.42 & 2014-03-29 & $> 16.95$     \\
Kiso & KWFC & $ g $ & 56747.72 & 2014-03-31 & $> 21.07$     \\
Kiso & KWFC & $ g $ & 56748.73 & 2014-04-01 & $> 20.00$     \\
Kiso & KWFC & $ g $ & 56749.74 & 2014-04-02 & $> 21.11$     \\
Kiso & KWFC & $ g $ & 56765.58 & 2014-04-18 & $> 19.63$     \\
Kiso & KWFC & $ g $ & 56766.49 & 2014-04-19 & $> 20.90$     \\
Kiso & KWFC & $ g $ & 56770.60 & 2014-04-23 & $21.21\pm0.20$      \\
Kiso & KWFC & $ g $ & 56771.57 & 2014-04-24 & $> 21.72$     \\
Kiso & KWFC & $ g $ & 56772.53 & 2014-04-25 & $> 21.32$     \\
Kiso & KWFC & $ g $ & 56773.54 & 2014-04-26 & $> 20.73$     \\
Kiso & KWFC & $ g $ & 56774.53 & 2014-04-27 & $> 21.72$     \\
Kiso & KWFC & $ g $ & 56809.54 & 2014-06-01 & $> 21.39$     \\
Kiso & KWFC & $ g $ & 56850.48 & 2014-07-12 & $> 18.33$     \\
Kiso & KWFC & $ g $ & 56853.49 & 2014-07-15 & $> 21.06$     \\
Kiso & KWFC & $ g $ & 56859.49 & 2014-07-21 & $> 20.43$     \\
Kiso & KWFC & $ g $ & 56876.47 & 2014-08-07 & $> 18.75$     \\
Kiso & KWFC & $ g $ & 56946.84 & 2014-10-16 & $> 18.70$     \\
Akeno & MITSuME & $ g $ & 56954.73 & 2014-10-24 & $> 19.87$     \\
Akeno & MITSuME & $ g $ & 56955.73 & 2014-10-25 & $> 20.75$     \\
Akeno & MITSuME & $ g $ & 56957.73 & 2014-10-27 & $> 19.09$     \\
OAO & MITSuME & $ g $ & 56957.80 & 2014-10-27 & $> 20.59$     \\
Kiso & KWFC & $ g $ & 56957.80 & 2014-10-27 & $> 20.57$     \\
Akeno & MITSuME & $ g $ & 56958.72 & 2014-10-28 & $> 20.72$     \\
Kiso & KWFC & $ g $ & 56958.78 & 2014-10-28 & $> 21.17$     \\
OAO & MITSuME & $ g $ & 56958.80 & 2014-10-28 & $> 20.06$     \\
Kiso & KWFC & $ g $ & 56959.78 & 2014-10-29 & $> 21.55$     \\
Murikabushi & MITSuME & $ g $ & 56959.80 & 2014-10-29 & $> 21.38$     \\
OAO & MITSuME & $ g $ & 56959.84 & 2014-10-29 & $> 19.20$     \\
Murikabushi & MITSuME & $ g $ & 56960.78 & 2014-10-30 & $21.24\pm0.22$      \\
Kiso & KWFC & $ g $ & 56960.80 & 2014-10-30 & $> 21.07$     \\
Murikabushi & MITSuME & $ g $ & 56961.78 & 2014-10-31 & $> 21.02$     \\
Akeno & MITSuME & $ g $ & 56962.73 & 2014-11-01 & $> 20.93$     \\
Akeno & MITSuME & $ g $ & 56963.71 & 2014-11-02 & $> 20.18$     \\
OAO & MITSuME & $ g $ & 56963.83 & 2014-11-02 & $> 19.36$     \\
Akeno & MITSuME & $ g $ & 56964.73 & 2014-11-03 & $> 20.70$     \\
Kiso & KWFC & $ g $ & 56964.79 & 2014-11-03 & $> 21.41$     \\
OAO & MITSuME & $ g $ & 56964.83 & 2014-11-03 & $> 20.31$     \\
Kiso & KWFC & $ g $ & 56965.76 & 2014-11-04 & $> 20.09$     \\
Akeno & MITSuME & $ g $ & 56965.78 & 2014-11-04 & $> 20.00$     \\
OAO & MITSuME & $ g $ & 56965.83 & 2014-11-04 & $> 20.14$     \\
Akeno & MITSuME & $ g $ & 56967.73 & 2014-11-06 & $> 19.89$     \\
Kiso & KWFC & $ g $ & 56967.79 & 2014-11-06 & $> 19.90$     \\
OAO & MITSuME & $ g $ & 56967.85 & 2014-11-06 & $> 18.55$     \\
Akeno & MITSuME & $ g $ & 56968.78 & 2014-11-07 & $> 20.07$     \\
Kiso & KWFC & $ g $ & 56968.82 & 2014-11-07 & $> 20.33$     \\
Kiso & KWFC & $ g $ & 56970.81 & 2014-11-09 & $> 20.67$     \\
OAO & MITSuME & $ g $ & 56971.83 & 2014-11-10 & $> 19.64$     \\
Akeno & MITSuME & $ g $ & 56973.70 & 2014-11-12 & $> 20.53$     \\
Akeno & MITSuME & $ g $ & 56974.69 & 2014-11-13 & $> 20.61$     \\
Akeno & MITSuME & $ g $ & 56975.68 & 2014-11-14 & $> 20.82$     \\
Akeno & MITSuME & $ g $ & 56976.68 & 2014-11-15 & $> 20.92$     \\
Kiso & KWFC & $ g $ & 56976.72 & 2014-11-15 & $> 20.60$     \\
Akeno & MITSuME & $ g $ & 56977.67 & 2014-11-16 & $> 20.64$     \\
Akeno & MITSuME & $ g $ & 56978.72 & 2014-11-17 & $> 20.56$     \\
Akeno & MITSuME & $ g $ & 56979.66 & 2014-11-18 & $> 21.07$     \\
Akeno & MITSuME & $ g $ & 56980.69 & 2014-11-19 & $> 20.00$     \\
Murikabushi & MITSuME & $ g $ & 56980.72 & 2014-11-19 & $> 20.55$     \\
Akeno & MITSuME & $ g $ & 56981.66 & 2014-11-20 & $> 21.00$     \\
Kiso & KWFC & $ g $ & 56981.69 & 2014-11-20 & $> 21.82$     \\
Murikabushi & MITSuME & $ g $ & 56981.73 & 2014-11-20 & $> 21.42$     \\
Kiso & KWFC & $ g $ & 56982.76 & 2014-11-21 & $> 21.15$     \\
Kiso & KWFC & $ g $ & 56983.74 & 2014-11-22 & $20.56\pm0.20$      \\
Nayuta & LISS & $ g $ & 56984.50 & 2014-11-23 & $> 20.19$     \\
Kiso & KWFC & $ g $ & 56984.72 & 2014-11-23 & $> 21.47$     \\
Kiso & KWFC & $ g $ & 56987.81 & 2014-11-26 & $> 21.44$     \\
Subaru & HSC & $ g $ & 56988.50 & 2014-11-27 & $21.72\pm0.12$      \\
Kiso & KWFC & $ g $ & 56988.67 & 2014-11-27 & $> 21.50$     \\
Murikabushi & MITSuME & $ g $ & 56989.71 & 2014-11-28 & $> 20.34$     \\
Kiso & KWFC & $ g $ & 57009.83 & 2014-12-18 & $19.51\pm0.12$      \\
Kiso & KWFC & $ g $ & 57014.60 & 2014-12-23 & $> 18.98$     \\
Kiso & KWFC & $ g $ & 57040.57 & 2015-01-18 & $> 19.53$     \\
Kiso & KWFC & $ g $ & 57042.60 & 2015-01-20 & $> 21.47$     \\
Kiso & KWFC & $ g $ & 57046.75 & 2015-01-24 & $> 21.53$     \\
Kiso & KWFC & $ g $ & 57047.77 & 2015-01-25 & $> 20.64$     \\
Kiso & KWFC & $ g $ & 57049.70 & 2015-01-27 & $> 21.45$     \\
Akeno & MITSuME & $ g $ & 57050.69 & 2015-01-28 & $> 20.85$     \\
Kiso & KWFC & $ g $ & 57055.72 & 2015-02-02 & $> 20.26$     \\
Akeno & MITSuME & $ g $ & 57062.54 & 2015-02-09 & $> 20.43$     \\
Akeno & MITSuME & $ g $ & 57063.67 & 2015-02-10 & $> 20.45$     \\
OAO & MITSuME & $ g $ & 57063.76 & 2015-02-10 & $> 19.19$     \\
OAO & MITSuME & $ g $ & 57064.61 & 2015-02-11 & $> 20.20$     \\
Akeno & MITSuME & $ g $ & 57064.74 & 2015-02-11 & $> 20.59$     \\
Kiso & KWFC & $ g $ & 57064.74 & 2015-02-11 & $> 20.91$     \\
Kiso & KWFC & $ g $ & 57073.63 & 2015-02-20 & $> 21.74$     \\
OAO188 & KOOLS & $ g $ & 57102.49 & 2015-03-21 & $19.82\pm0.14$      \\
OAO188 & KOOLS & $ g $ & 57103.46 & 2015-03-22 & $18.70\pm0.11$      \\
OAO188 & KOOLS & $ g $ & 57103.56 & 2015-03-22 & $18.58\pm0.11$      \\
OAO188 & KOOLS & $ g $ & 57103.72 & 2015-03-22 & $18.94\pm0.10$      \\
OAO188 & KOOLS & $ g $ & 57104.56 & 2015-03-23 & $19.45\pm0.12$      \\
OAO & MITSuME & $ g $ & 57106.64 & 2015-03-25 & $> 20.61$     \\
OAO & MITSuME & $ g $ & 57129.59 & 2015-04-17 & $> 20.90$     \\
OAO & MITSuME & $ g $ & 57133.60 & 2015-04-21 & $> 21.04$     \\
OAO & MITSuME & $ g $ & 57166.57 & 2015-05-24 & $> 20.59$     \\
Subaru & HSC & $ g $ & 57167.50 & 2015-05-25 & $21.74\pm0.13$      \\
OAO & MITSuME & $ g $ & 57167.53 & 2015-05-25 & $> 19.49$     \\
OAO & MITSuME & $ g $ & 57174.49 & 2015-06-01 & $> 19.14$     \\
OAO & MITSuME & $ g $ & 57176.54 & 2015-06-03 & $> 20.18$     \\
\hline
Kottamia & - & $ V $ & 56726.93 & 2014-03-10 & $19.17\pm0.13$      \\
Saitama & - & $ V $ & 57063.54 & 2015-02-10 & $> 18.92$     \\
\hline
Kanata & HONIR & $ R_C $ & 56712.62 & 2014-02-24 & $20.48\pm0.16$      \\
Akeno & MITSuME & $ R_C $ & 56726.51 & 2014-03-10 & $19.48\pm0.21$      \\
Kanata & HONIR & $ R_C $ & 56726.75 & 2014-03-10 & $19.73\pm0.12$      \\
Kottamia & - & $ R_C $ & 56726.93 & 2014-03-10 & $20.01\pm0.14$      \\
Akeno & MITSuME & $ R_C $ & 56954.73 & 2014-10-24 & $> 19.56$     \\
Akeno & MITSuME & $ R_C $ & 56955.73 & 2014-10-25 & $20.04\pm0.19$      \\
Akeno & MITSuME & $ R_C $ & 56957.73 & 2014-10-27 & $> 18.16$     \\
OAO & MITSuME & $ R_C $ & 56957.80 & 2014-10-27 & $> 19.95$     \\
Akeno & MITSuME & $ R_C $ & 56958.72 & 2014-10-28 & $20.24\pm0.20$      \\
OAO & MITSuME & $ R_C $ & 56958.80 & 2014-10-28 & $> 19.75$     \\
Murikabushi & MITSuME & $ R_C $ & 56959.80 & 2014-10-29 & $20.35\pm0.18$      \\
OAO & MITSuME & $ R_C $ & 56959.84 & 2014-10-29 & $> 19.03$     \\
Murikabushi & MITSuME & $ R_C $ & 56960.78 & 2014-10-30 & $20.51\pm0.19$      \\
Murikabushi & MITSuME & $ R_C $ & 56961.78 & 2014-10-31 & $20.73\pm0.18$      \\
Akeno & MITSuME & $ R_C $ & 56962.73 & 2014-11-01 & $> 19.97$     \\
Akeno & MITSuME & $ R_C $ & 56963.71 & 2014-11-02 & $> 19.68$     \\
OAO & MITSuME & $ R_C $ & 56963.83 & 2014-11-02 & $> 19.13$     \\
Akeno & MITSuME & $ R_C $ & 56964.73 & 2014-11-03 & $> 20.14$     \\
OAO & MITSuME & $ R_C $ & 56964.83 & 2014-11-03 & $> 19.65$     \\
Akeno & MITSuME & $ R_C $ & 56965.75 & 2014-11-04 & $> 20.13$     \\
OAO & MITSuME & $ R_C $ & 56965.83 & 2014-11-04 & $> 19.69$     \\
Akeno & MITSuME & $ R_C $ & 56967.74 & 2014-11-06 & $> 19.94$     \\
OAO & MITSuME & $ R_C $ & 56967.85 & 2014-11-06 & $> 18.32$     \\
Akeno & MITSuME & $ R_C $ & 56968.78 & 2014-11-07 & $> 19.60$     \\
OAO & MITSuME & $ R_C $ & 56971.83 & 2014-11-10 & $> 19.11$     \\
Akeno & MITSuME & $ R_C $ & 56973.70 & 2014-11-12 & $> 19.89$     \\
Akeno & MITSuME & $ R_C $ & 56974.69 & 2014-11-13 & $> 20.37$     \\
Akeno & MITSuME & $ R_C $ & 56975.68 & 2014-11-14 & $> 19.91$     \\
Akeno & MITSuME & $ R_C $ & 56976.68 & 2014-11-15 & $20.14\pm0.19$      \\
Kanata & HONIR & $ R_C $ & 56976.87 & 2014-11-15 & $> 18.18$     \\
Kanata & HONIR & $ R_C $ & 56976.87 & 2014-11-15 & $> 19.67$     \\
Akeno & MITSuME & $ R_C $ & 56977.67 & 2014-11-16 & $19.58\pm0.19$      \\
Akeno & MITSuME & $ R_C $ & 56978.71 & 2014-11-17 & $> 20.03$     \\
Akeno & MITSuME & $ R_C $ & 56979.66 & 2014-11-18 & $> 20.50$     \\
Akeno & MITSuME & $ R_C $ & 56980.69 & 2014-11-19 & $> 19.45$     \\
Murikabushi & MITSuME & $ R_C $ & 56980.72 & 2014-11-19 & $> 20.32$     \\
Akeno & MITSuME & $ R_C $ & 56981.66 & 2014-11-20 & $> 20.15$     \\
Murikabushi & MITSuME & $ R_C $ & 56981.73 & 2014-11-20 & $20.86\pm0.18$      \\
Kanata & HONIR & $ R_C $ & 56981.73 & 2014-11-20 & $21.01\pm0.13$      \\
Murikabushi & MITSuME & $ R_C $ & 56989.71 & 2014-11-28 & $> 20.04$     \\
Akeno & MITSuME & $ R_C $ & 57050.69 & 2015-01-28 & $> 20.18$     \\
Akeno & MITSuME & $ R_C $ & 57062.54 & 2015-02-09 & $> 19.79$     \\
Akeno & MITSuME & $ R_C $ & 57063.67 & 2015-02-10 & $> 20.26$     \\
OAO & MITSuME & $ R_C $ & 57063.76 & 2015-02-10 & $> 18.85$     \\
OAO & MITSuME & $ R_C $ & 57064.61 & 2015-02-11 & $> 19.74$     \\
Akeno & MITSuME & $ R_C $ & 57064.74 & 2015-02-11 & $> 20.24$     \\
OAO188 & KOOLS & $ R_C $ & 57102.50 & 2015-03-21 & $19.27\pm0.12$      \\
OAO188 & KOOLS & $ R_C $ & 57102.67 & 2015-03-21 & $18.95\pm0.11$      \\
OAO188 & KOOLS & $ R_C $ & 57102.80 & 2015-03-21 & $18.62\pm0.11$      \\
OAO188 & KOOLS & $ R_C $ & 57103.48 & 2015-03-22 & $18.13\pm0.10$      \\
OAO188 & KOOLS & $ R_C $ & 57103.58 & 2015-03-22 & $18.07\pm0.10$      \\
OAO & MITSuME & $ R_C $ & 57106.64 & 2015-03-25 & $20.35\pm0.20$      \\
Kanata & HONIR & $ R_C $ & 57106.66 & 2015-03-25 & $20.64\pm0.12$      \\
OAO & MITSuME & $ R_C $ & 57129.59 & 2015-04-17 & $19.48\pm0.18$      \\
Akeno & MITSuME & $ R_C $ & 57130.66 & 2015-04-18 & $> 20.07$     \\
OAO & MITSuME & $ R_C $ & 57133.60 & 2015-04-21 & $19.98\pm0.17$      \\
OAO & MITSuME & $ R_C $ & 57166.57 & 2015-05-24 & $> 19.98$     \\
OAO & MITSuME & $ R_C $ & 57167.53 & 2015-05-25 & $> 19.13$     \\
OAO & MITSuME & $ R_C $ & 57174.49 & 2015-06-01 & $> 18.88$     \\
OAO & MITSuME & $ R_C $ & 57176.54 & 2015-06-03 & $> 19.87$     \\
\hline
Subaru & FOCAS & $ i $ & 56712.63 & 2014-02-24 & $20.72\pm0.10$      \\
\hline
Kanata & HONIR & $ I_C $ & 56726.77 & 2014-03-10 & $19.41\pm0.14$      \\
Kottamia & - & $ I_C $ & 56732.53 & 2014-03-16 & $18.91\pm0.21$      \\
Akeno & MITSuME & $ I_C $ & 56954.73 & 2014-10-24 & $> 18.85$     \\
Akeno & MITSuME & $ I_C $ & 56955.73 & 2014-10-25 & $> 19.68$     \\
Akeno & MITSuME & $ I_C $ & 56957.73 & 2014-10-27 & $> 17.19$     \\
OAO & MITSuME & $ I_C $ & 56957.80 & 2014-10-27 & $> 19.37$     \\
Akeno & MITSuME & $ I_C $ & 56958.72 & 2014-10-28 & $> 19.64$     \\
OAO & MITSuME & $ I_C $ & 56958.80 & 2014-10-28 & $> 19.29$     \\
Murikabushi & MITSuME & $ I_C $ & 56959.80 & 2014-10-29 & $19.99\pm0.19$      \\
OAO & MITSuME & $ I_C $ & 56959.84 & 2014-10-29 & $> 18.32$     \\
Murikabushi & MITSuME & $ I_C $ & 56960.78 & 2014-10-30 & $20.08\pm0.19$      \\
Murikabushi & MITSuME & $ I_C $ & 56961.78 & 2014-10-31 & $> 19.56$     \\
Akeno & MITSuME & $ I_C $ & 56962.73 & 2014-11-01 & $> 19.51$     \\
Akeno & MITSuME & $ I_C $ & 56963.71 & 2014-11-02 & $> 19.11$     \\
OAO & MITSuME & $ I_C $ & 56963.83 & 2014-11-02 & $> 18.43$     \\
Akeno & MITSuME & $ I_C $ & 56964.71 & 2014-11-03 & $> 19.65$     \\
OAO & MITSuME & $ I_C $ & 56964.83 & 2014-11-03 & $> 19.21$     \\
Akeno & MITSuME & $ I_C $ & 56965.72 & 2014-11-04 & $> 19.68$     \\
OAO & MITSuME & $ I_C $ & 56965.83 & 2014-11-04 & $> 19.19$     \\
Akeno & MITSuME & $ I_C $ & 56967.72 & 2014-11-06 & $> 19.55$     \\
OAO & MITSuME & $ I_C $ & 56967.85 & 2014-11-06 & $> 17.92$     \\
Akeno & MITSuME & $ I_C $ & 56968.77 & 2014-11-07 & $> 19.19$     \\
OAO & MITSuME & $ I_C $ & 56971.83 & 2014-11-10 & $> 18.80$     \\
Akeno & MITSuME & $ I_C $ & 56973.70 & 2014-11-12 & $> 19.27$     \\
Akeno & MITSuME & $ I_C $ & 56974.69 & 2014-11-13 & $> 19.75$     \\
Akeno & MITSuME & $ I_C $ & 56975.68 & 2014-11-14 & $> 19.61$     \\
Akeno & MITSuME & $ I_C $ & 56976.68 & 2014-11-15 & $> 19.76$     \\
Akeno & MITSuME & $ I_C $ & 56977.67 & 2014-11-16 & $> 19.65$     \\
Akeno & MITSuME & $ I_C $ & 56978.71 & 2014-11-17 & $> 19.61$     \\
Akeno & MITSuME & $ I_C $ & 56979.66 & 2014-11-18 & $> 19.88$     \\
Akeno & MITSuME & $ I_C $ & 56980.69 & 2014-11-19 & $> 18.84$     \\
Murikabushi & MITSuME & $ I_C $ & 56980.72 & 2014-11-19 & $> 19.94$     \\
Akeno & MITSuME & $ I_C $ & 56981.66 & 2014-11-20 & $> 19.63$     \\
Murikabushi & MITSuME & $ I_C $ & 56981.73 & 2014-11-20 & $> 20.70$     \\
Murikabushi & MITSuME & $ I_C $ & 56989.71 & 2014-11-28 & $> 19.02$     \\
Akeno & MITSuME & $ I_C $ & 57050.65 & 2015-01-28 & $> 19.78$     \\
Akeno & MITSuME & $ I_C $ & 57062.54 & 2015-02-09 & $> 18.27$     \\
Akeno & MITSuME & $ I_C $ & 57063.67 & 2015-02-10 & $> 19.73$     \\
OAO & MITSuME & $ I_C $ & 57063.76 & 2015-02-10 & $> 18.36$     \\
OAO & MITSuME & $ I_C $ & 57064.61 & 2015-02-11 & $> 18.83$     \\
Akeno & MITSuME & $ I_C $ & 57064.74 & 2015-02-11 & $> 19.69$     \\
OAO188 & KOOLS & $ I_C $ & 57102.49 & 2015-03-21 & $19.07\pm0.12$      \\
OAO188 & KOOLS & $ I_C $ & 57102.66 & 2015-03-21 & $18.86\pm0.11$      \\
OAO188 & KOOLS & $ I_C $ & 57102.79 & 2015-03-21 & $18.53\pm0.11$      \\
OAO188 & KOOLS & $ I_C $ & 57103.47 & 2015-03-22 & $18.28\pm0.11$      \\
OAO188 & KOOLS & $ I_C $ & 57103.57 & 2015-03-22 & $17.92\pm0.10$      \\
OAO & MITSuME & $ I_C $ & 57106.64 & 2015-03-25 & $> 19.79$     \\
OAO & MITSuME & $ I_C $ & 57129.59 & 2015-04-17 & $> 19.60$     \\
Akeno & MITSuME & $ I_C $ & 57130.66 & 2015-04-18 & $> 19.38$     \\
OAO & MITSuME & $ I_C $ & 57133.60 & 2015-04-21 & $19.55\pm0.21$      \\
OAO & MITSuME & $ I_C $ & 57166.57 & 2015-05-24 & $> 19.44$     \\
OAO & MITSuME & $ I_C $ & 57167.53 & 2015-05-25 & $> 18.72$     \\
OAO & MITSuME & $ I_C $ & 57174.49 & 2015-06-01 & $> 18.45$     \\
OAO & MITSuME & $ I_C $ & 57176.54 & 2015-06-03 & $> 19.57$     \\
\hline
Kiso & KWFC & $ z $ & 56958.82 & 2014-10-28 & $> 19.78$     \\
Kiso & KWFC & $ z $ & 56959.84 & 2014-10-29 & $> 19.70$     \\
Nayuta & LISS & $ z $ & 56984.50 & 2014-11-23 & $19.51\pm0.12$      \\
Subaru & HSC & $ z $ & 56987.50 & 2014-11-26 & $20.57\pm0.11$      \\
Kiso & KWFC & $ z $ & 57064.76 & 2015-02-11 & $> 18.99$     \\
OAO188 & KOOLS & $ z $ & 57102.67 & 2015-03-21 & $18.79\pm0.12$      \\
OAO188 & KOOLS & $ z $ & 57103.47 & 2015-03-22 & $17.96\pm0.12$      \\
OAO188 & KOOLS & $ z $ & 57103.58 & 2015-03-22 & $17.78\pm0.11$      \\
\hline
Kanata & HONIR & $ J $ & 56732.68 & 2014-03-16 & $18.32\pm0.21$      \\
Nayuta & NIC & $ J $ & 56981.75 & 2014-11-20 & $20.23\pm0.21$      \\
Nayuta & NIC & $ J $ & 57104.53 & 2015-03-23 & $18.38\pm0.20$      \\
Nayuta & NIC & $ J $ & 57106.67 & 2015-03-25 & $19.85\pm0.21$      \\
\hline
Nayuta & NIC & $ H $ & 56955.80 & 2014-10-25 & $19.64\pm0.21$      \\
Nayuta & NIC & $ H $ & 56964.79 & 2014-11-03 & $19.60\pm0.21$      \\
Nayuta & NIC & $ H $ & 56965.77 & 2014-11-04 & $19.62\pm0.21$      \\
Nayuta & NIC & $ H $ & 56981.76 & 2014-11-20 & $20.01\pm0.20$      \\
Nayuta & NIC & $ H $ & 57104.53 & 2015-03-23 & $17.79\pm0.20$      \\
\hline
Kanata & HONIR & $ K_{s} $ & 56726.77 & 2014-03-10 & $17.85\pm0.21$      \\
Kanata & HONIR & $ K_{s} $ & 56732.70 & 2014-03-16 & $17.68\pm0.22$      \\
Kagoshima & - & $ K_{s} $ & 56954.77 & 2014-10-24 & $> 14.87$     \\
Nayuta & NIC & $ K_{s} $ & 56955.80 & 2014-10-25 & $18.91\pm0.21$      \\

OAO188 & ISLE & $ K $ & 56959.86 & 2014-10-29 & $19.02\pm0.21$      \\

Nayuta & NIC & $ K_{s} $ & 56964.79 & 2014-11-03 & $> 18.89$     \\
Kagoshima & - & $ K_{s} $ & 56965.74 & 2014-11-04 & $> 16.85$     \\
Nayuta & NIC & $ K_{s} $ & 56965.77 & 2014-11-04 & $19.49\pm0.21$      \\
Kagoshima & - & $ K_{s} $ & 56975.71 & 2014-11-14 & $> 15.85$     \\
Kanata & HONIR & $ K_{s} $ & 56976.87 & 2014-11-15 & $> 16.75$     \\
Kagoshima & - & $ K_{s} $ & 56981.71 & 2014-11-20 & $> 16.15$     \\
Kanata & HONIR & $ K_{s} $ & 56981.72 & 2014-11-20 & $> 17.27$     \\
Nayuta & NIC & $ K_{s} $ & 56981.76 & 2014-11-20 & $19.73\pm0.21$      \\

OAO188 & ISLE & $ K $ & 56981.84 & 2014-11-20 & $19.11\pm0.20$      \\

Nayuta & NIC & $ K_{s} $ & 56982.86 & 2014-11-21 & $> 18.89$     \\
Kagoshima & - & $ K_{s} $ & 56988.70 & 2014-11-27 & $> 14.47$     \\
Nayuta & NIC & $ K_{s} $ & 57104.51 & 2015-03-23 & $17.45\pm0.20$      \\
Kanata & HONIR & $ K_{s} $ & 57106.66 & 2015-03-25 & $> 17.75$     \\
Nayuta & NIC & $ K_{s} $ & 57106.67 & 2015-03-25 & $19.22\pm0.20$      \\